\documentclass[journal]{IEEEtran}
\usepackage{graphicx}
\graphicspath{{./images/}}
\usepackage[justification=centering]{caption}
\usepackage{amsmath}
\usepackage{amssymb}
\usepackage{booktabs}
\usepackage{float}
\usepackage{tabularx}

\usepackage{colortbl}
\usepackage{xcolor}

\ifCLASSINFOpdf
\else
\fi
\usepackage{algorithm}
\usepackage{algpseudocode}

\begin{document}

\title{PECL: A Heterogeneous Parallel Multi-Domain Network for Radar-Based Human Activity Recognition}

\author{Jiuqi~Yan,
        Chendong~Xu,
        Dongyu~Liu,
}


\maketitle

\begin{abstract}
Radar systems are increasingly favored for medical applications because they provide non-intrusive monitoring with high privacy and robustness to lighting conditions. However, existing research typically relies on single-domain radar signals and overlooks the temporal dependencies inherent in human activity, which complicates the classification of similar actions. To address this issue, we designed the Parallel-EfficientNet-CBAM-LSTM (PECL) network to process data in three complementary domains: Range-Time, Doppler-Time, and Range-Doppler. PECL combines a channel-spatial attention module and temporal units to capture more features and dynamic dependencies during action sequences, improving both accuracy and robustness. The experimental results show that PECL achieves an accuracy of 96.16\% on the same dataset, outperforming existing methods by at least 4.78\%. PECL also performs best in distinguishing between easily confused actions. Despite its strong performance, PECL maintains moderate model complexity, with 23.42M parameters and 1324.82M FLOPs. Its parameter-efficient design further reduces computational cost. 
\end{abstract}

\begin{IEEEkeywords}
Human Activity Recognition, Radar Systems, Multi-Domain Representation, Deep Learning, Attention Module, Computational Costs. 
\end{IEEEkeywords}

\IEEEpeerreviewmaketitle

\section{Introduction}

Human activity recognition (HAR) has long been an active research area. With the acceleration of population aging, demand for HAR technology is growing in both hospitals and households \cite{florence2018medical}\cite{key}\cite{nicholson2024interventions}. 

In recent years, research on HAR using different types of sensors has grown \cite{habibzadeh2019survey}. Based on sensor type, HAR can be categorized into contact-based and non-contact-based. Contact-base HAR sensors include pressure sensors \cite{agrawal2023fall}\cite{youngkong2021novel}\cite{brome2022fall}, gyroscopes \cite{mukherjee2023novel}\cite{paramasivam2022internet}\cite{matos2023sensitivity}, and accelerometers \cite{boudouane2023portable}\cite{lahouar2023development}\cite{abd2018fall}. These sensors offer high detection accuracy and are relatively low-cost. However, they require prolonged use, which compromises user comfort. In many scenarios, users cannot maintain continuous wear, creating potential safety risks during non-wearing periods. Over the past few years, most non-contact HAR systems have relied on visual cameras \cite{ros2023flexible}\cite{pranavan2023fall}. With the aid of deep learning, camera-based HAR has achieved high accuracy. However, cameras are susceptible to lighting conditions and occlusions, and they raise significant privacy concerns. Many falls occur in bathrooms, which are private spaces where users typically do not install cameras. Additionally, steam can reduce camera visibility, affecting recognition accuracy. 

Millimeter-wave radar-based detection solutions effectively address these shortcomings. Millimeter-wave radar is a non-contact sensor that is unaffected by lighting conditions. Radar does not record optical images or videos that could expose privacy. In the last decade, there have been many studies on radar-based HAR \cite{li2023radar}\cite{lu2021radar}\cite{ullmann2023survey}\cite{sadreazami2019fall}\cite{wang2020millimetre}\cite{abd2023passive}\cite{cao2024real}\cite{yao2023unsupervised}\cite{chen2022three}. However, there are still many shortcomings and challenges in research on radar-based HAR. Most studies focus only on information from a single domain, and the lack of rich input information leads to insufficient accuracy in activity recognition. The time-related nature of each movement is also often ignored. Each activity label encompasses a sequence of subtle motions, which are key to distinguishing between easily confused movement categories. 

In this paper, we propose a parallel-structured network: Parallel-EfficientNet-CBAM-LSTM (PECL). PECL can simultaneously process radar data from three complementary domains: Range-Time domain, Doppler-Time domain, and Range-Doppler domain. PECL can simultaneously capture spatial and time information, providing more complete information for action recognition. The three branches of PECL are modified based on the EfficientNet architecture, replacing the original Squeeze-and-Excitation (SE)\cite{hu2018squeeze} module with the Convolutional Block Attention Module (CBAM)\cite{woo2018cbam}. There is “spatial information” in the spectrograms, and each area of the diagram may contain critical information. CBAM's spatial attention gives these spatial locations higher semantic weight. LSTM modules are added to the Range-Time and Doppler-Time branches to capture the temporal correlation of motion. These structures help PECL significantly reduce misclassification of similar actions. Ultimately, PECL achieved an accuracy rate of 96.16\% on dataset \cite{fioranelli2019radar2}. On the same dataset, PECL improved at least 4.78\% compared with other baseline methods. In the dataset, “Pick Up” and “Drink” are the two action categories that are most easily confused with each other. PECL achieved 89.77\% and 90.38\% accuracy in identifying 'Pick Up' and 'Drink', respectively, achieving the best performance among all compared models. When verifying generalization capabilities using different datasets, PECL also achieved the best results.  
 
Section II presents and analyzes recent relevant studies. In Section III, we introduce the modeling of radar signals and how to obtain information in the Range-Time, Doppler-Time, and Range-Doppler domains. We also introduce the PECL proposed in this paper. In Section IV, we present experimental details, including data source, noise addition, parameters of our network, and training details. We present the experimental results in Section V and compare them with existing studies. We use ablation experiments to illustrate the significance of each structure and module in the model. In Section VI, we summarize the experiments and outline directions for future work. 


\section{Ralated Works}
Recent studies have demonstrated the potential of radar-based systems for human activity recognition due to their robustness to lighting and privacy-preserving properties.

C. Ding et al.~\cite{ding2021fall} proposed a fall detection method based on MIMO millimeter-wave radar and KNN, which achieves high accuracy while ensuring low complexity by manually extracting velocity, acceleration, and DOA change rate features. 

M. Chen et al.~\cite{chen2022three} proposed a three-stage low-complexity detection framework. Large motion detection: Detect sparse fall events through high-frequency Doppler energy detection. Coarse fall detection: Extract six time-frequency features and two position features, and use SVDD (Support Vector Data Description) to distinguish between non-fall and fall-like events (such as rapid squatting). Enhanced fall detection: Adds a trunk average velocity feature and combines it with a Mahalanobis distance classifier for final fall determination. SVDD is trained using a large number of non-fall samples to reduce the difficulty of direct classification. Mahalanobis distance improves accuracy by reducing the search space. 

Zhang et al.~\cite{zhang2020phase} proposed a coherent integration method known as phase compensation transformation (PCT) to enhance the detection capability of linear FMCW radar for human targets. By modeling human activity using the Boulic model, they constructed phase compensation signals to correct nonlinear phase shifts caused by micro-Doppler effects, enabling more effective signal accumulation via FFT.

Li et al.~\cite{li2023radar} investigated human activity recognition using radar signals with a focus on adaptive thresholding. Their method highlighted regions of interest in spectrograms and combined multiple domains—such as masked spectrograms, raw phase, and unwrapped phase—to improve classification accuracy.

Lu et al.~\cite{lu2021radar} developed a low-power Doppler radar system for real-time fall detection. Their multi-stage approach included event detection, fall-like event filtering, and final fall classification. To reduce power consumption, deep learning modules were activated only when potential fall events were detected.

Yao et al.~\cite{yao2023unsupervised} proposed an unsupervised learning method for fall detection using only non-fall samples. They extracted temporal, spatial, and velocity features via 3D convolutional and transposed convolutional layers, and trained a predictor to model the patterns of normal activity. To improve robustness, hard sample mining (HSM) was employed to refine model decision boundaries.

X. Yao~\cite{yao2020human} introduced a complex-valued convolutional neural network (CV-CNN) for radar-based human activity recognition. Unlike traditional real-valued CNNs, their model utilized both amplitude and phase information from the complex radar signals, resulting in better learning of micro-Doppler features and improved classification performance.

Janakaraj et al.~\cite{janakaraj2019star} proposed STAR, a mmWave radar-based gait recognition system designed to simultaneously track human targets and identify individuals. The system included both tracking and identity recognition modules, supporting real-time biometric analysis based on motion signatures.

He et al.~\cite{he2023fall} considered multiple time-related radar representations, such as Range-Time, Doppler-Time, and Angle-Time maps, for fall detection. However, their method did not utilize the Range-Doppler domain, which offers complementary spatial-frequency information that could enhance the recognition of dynamic motion.

While the above research has contributed significantly to radar-based activity recognition, most of them either focus on single-domain information or treat temporal dynamics in a limited fashion. Human actions are inherently continuous and composed of a series of temporally correlated micro-movements. Ignoring such continuity may lead to suboptimal recognition performance. Furthermore, the integration of spatial and temporal features from multiple domains remains underexplored.

To address these challenges, we propose a multi-domain learning framework that simultaneously leverages Range-Time, Doppler-Time, and Range-Doppler representations. Combined with temporal modeling via LSTM and spatial attention through CBAM, our approach aims to fully exploit the rich information encoded in radar echoes for more accurate and robust human activity recognition. 

\section{Methodology}
\begin{figure}[htb]
    \centering
    \includegraphics[scale=0.315]{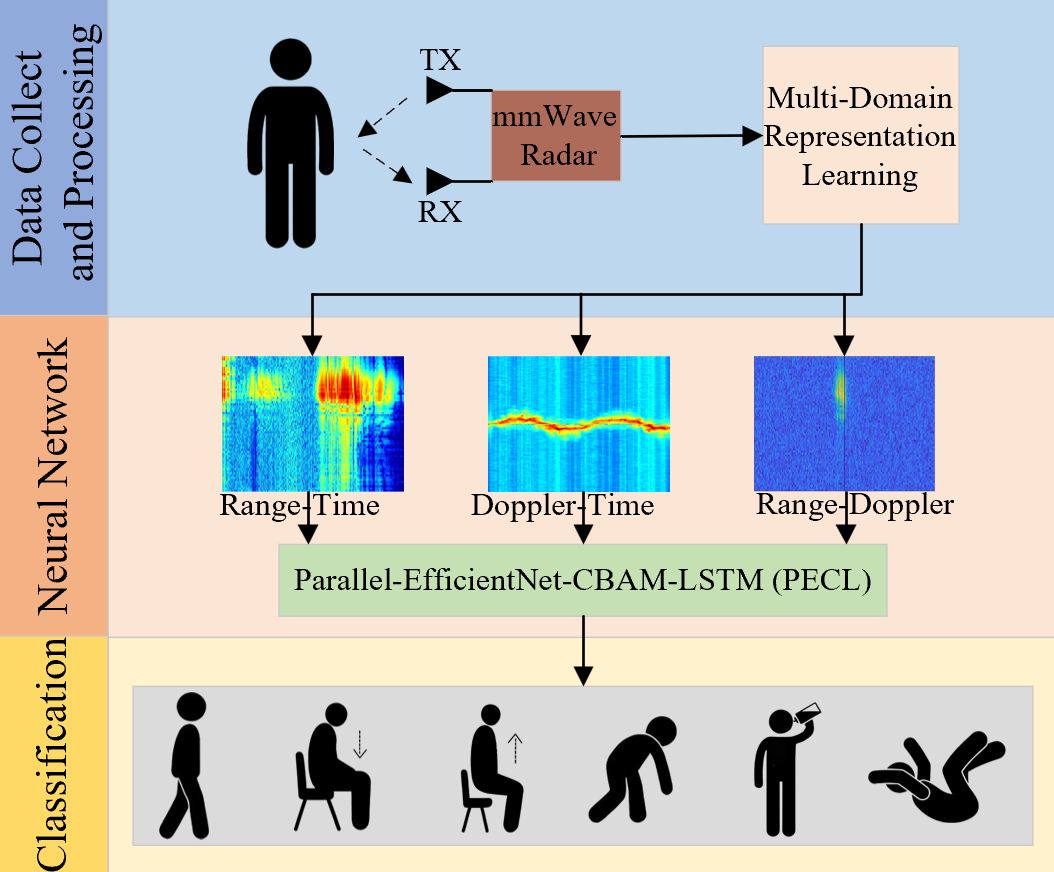}
    \caption{System flowchart of the radar HAR pipeline}
    \label{Progress}
\end{figure}

The HAR system process consists of three steps: radar echo signal processing, neural network, and classification. The system flowchart is shown in Figure~\ref{Progress}. 

\subsection{Echo Signal Model of Radar}
The IF (Intermediate Frequency) signal can be modeled as
\begin{equation}
s_b(t) = \cos \left[ 2\pi f_b t + \phi(t) \right], \quad f_b = \frac{2kR}{c},
\end{equation}
where $f_b$ is the beat frequency encoding target range, $k = B/T$ is the chirp slope with bandwidth $B$ and duration $T$, $R$ is the target range, and $c$ is the speed of light. The phase term $\phi(t)$ captures Doppler effects and is expressed as
\begin{equation}
\phi(t) = \frac{4\pi R(t)}{\lambda},
\end{equation}
where $R(t)$ is the instantaneous target range and $\lambda$ is the radar wavelength. 

For digital processing, the continuous IF signal $s_b(t)$ is sampled at rate $f_s$. Each chirp yields $N_s = T \cdot f_s$ fast-time samples, and a total of $N_c$ chirps form one measurement frame. The sampled data are arranged into a two-dimensional matrix
\begin{equation}
s(n,m) \in \mathbb{R}^{N_c \times N_s},
\end{equation}
where $n$ indexes chirps (slow time) and $m$ indexes ADC samples within each chirp (fast time). 

\subsection{Multi-Domain Representation Learning}

\subsubsection{Range-Time Domain}

The Range-Time map (RTM) characterizes how target range evolves across successive chirps. Starting from the sampled data $s[n,m] \in \mathbb{R}^{N_c \times N_s}$, a one-dimensional discrete Fourier transform (DFT) is applied along the fast-time axis to obtain range profiles:
\begin{equation}
S_{\text{RT}}[n, r] = \sum_{m=0}^{N_s - 1} s[n, m] \cdot w[m] \cdot e^{-j \frac{2\pi}{N_s} r m},
\end{equation}
where $w[m]$ is a window function (rectangular in our implementation) and $r$ denotes the range bin index. (Practically, the DFT is computed via the fast Fourier transform (FFT).)

To suppress stationary clutter, a moving target indication (MTI) filter is applied along the slow-time dimension using a fourth-order Butterworth high-pass filter with a normalized cutoff frequency of 0.0075 (relative to the Nyquist frequency): 
\begin{equation}
S_{\text{RT-MTI}}[n, r] = \sum_{k=0}^{4} b_k \, x[n-k] - \sum_{k=1}^{4} a_k \, y[n-k],
\end{equation}
where $x[n] = |S_{\text{RT}}[n,r]|$ is the magnitude sequence at range bin $r$, $y[n]$ is the filtered output, and $\{a_k, b_k\}$ are the Butterworth high-pass filter. Here, $k$ indexes the filter taps.

Finally, the RTM is visualized using a log-magnitude transformation:
\begin{equation}
\mathcal{R}[n, r] = 20 \log_{10} \left( |S_{\text{RT-MTI}}[n, r]| \right),
\end{equation}
which enhances weak motion signatures and compresses the dynamic range, 
allowing clearer visualization of moving targets while attenuating static reflections. 

\subsubsection{Doppler-Time Domain}

The Doppler-Time map (DTM) reveals velocity variations over time and is particularly sensitive to periodic movements such as 'walking' or 'falling'. For each selected range bin \( r \in [r_1, r_2] \), we extract the slow-time signal \( s_r[n] \) and apply a Adaptive-Short-Time Fourier transform (ASTFT) with a sliding window: 
\begin{equation}
\operatorname{ASTFT}_r[n, f_D] = \sum_{k=0}^{N_c - 1} s_r[k] \cdot w[k - n] \cdot e^{-j \frac{2\pi}{N_c} f_D k},
\end{equation}
where \( w[\cdot] \) is the analysis window whose shape is adaptively selected at each time index based on energy concentration, and \( f_D \) is the Doppler bin index. 

Unlike fixed-window STFT, ASTFT dynamically selects window widths to improve time-frequency localization, especially for signals with nonstationary features. 
Concretely, a bank of Gaussian windows with varying shape parameters $\alpha$ is evaluated, and at each time index $n$ the window is chosen by minimizing a spectral concentration factor:
\begin{equation}
P(n) = \arg\min_{\alpha} \; \frac{\left(\sum_{f} |X_{\alpha}(n,f)| \right)^2}{\sum_{f} |X_{\alpha}(n,f)|^2 + \epsilon},
\end{equation}
where $X_{\alpha}(n,f)$ denotes the short-time spectrum at time index $n$ and frequency bin $f$ using a Gaussian window with parameter $\alpha$, and $\epsilon$ is a small constant for numerical stability.

This adaptive selection allows narrow windows to be used for transients and wider windows for quasi-stationary segments, thus enhancing time-frequency resolution. 

The final DTM is obtained by aggregating the magnitude spectra over a contiguous set of range bins:
\begin{equation}
\mathcal{D}[n, f_D] = \sum_{r = r_1}^{r_2} \left| \operatorname{ASTFT}_r[n, f_D] \right|,
\end{equation}
where $r_1$ and $r_2$ specify the lower and upper bounds of the selected range-bin interval. 

This representation shows the relationship between the speed of the moving target and time, allowing us to clearly distinguish the details of different actions. 

\subsubsection{Range-Doppler Domain}

After we get \( S_{\text{RT-MTI}}[n, r]\), a second DFT is applied along the slow-time dimension to extract Doppler signatures:  
\begin{equation}
S_{\text{RD}}[r, f_D] = \sum_{n=0}^{N_c - 1} S_{\text{RT-MTI}}[n, r] \cdot w[n] \cdot e^{-j \frac{2\pi}{N_c} f_D n},
\end{equation}
where \( w[n] \) is a rectangular window. (Again, this DFT is implemented using FFT.)

The final RDM is obtained via log-magnitude transformation:
\begin{equation}
\mathcal{M}[r, f_D] = 20 \log_{10} \left( \left| S_{\text{RD}}[r, f_D] \right| \right).
\end{equation}

This 2D representation effectively captures motion patterns across both range and Doppler domains and is well-suited for distinguishing stationary and dynamic activities.

\subsection{Parallel-EfficientNet-CBAM-LSTM (PECL)}
\subsubsection{Structure of PECL}
For network design, we employ the Parallel-EfficientNet-CBAM-LSTM (PECL) architecture, which consists of three parallel EfficientNet-B0 branches. EfficientNet-B0 is chosen for its excellent balance between accuracy and computational efficiency. When pretrained on ImageNet, this network contains 5.33M parameters (5.29M trainable). 

The PECL network processes three distinct input modalities in parallel: Range-Time Maps, Doppler-Time Maps, and Range-Doppler Maps, with each modality handled by a dedicated branch. All three branches share a similar structure comprising a 3×3 convolutional layer, MBConv modules with CBAM attention mechanism, and a 1×1 convolutional layer. The key architectural differences lie in the final processing stages: both the Range-Time and Doppler-Time branches incorporate LSTM modules to capture temporal dynamics, while the Range-Doppler branch employs a linear layer followed by maximum pooling. Each of the three branches outputs a 128-dimensional feature vector. These vectors are concatenated to form a unified 384-dimensional feature representation. Prior to feeding this fused vector into the final fully-connected layer, we apply a dropout layer with a rate of 0.2 to prevent overfitting. 

\begin{figure}[htb]
    \centering
    \includegraphics[scale=0.75]{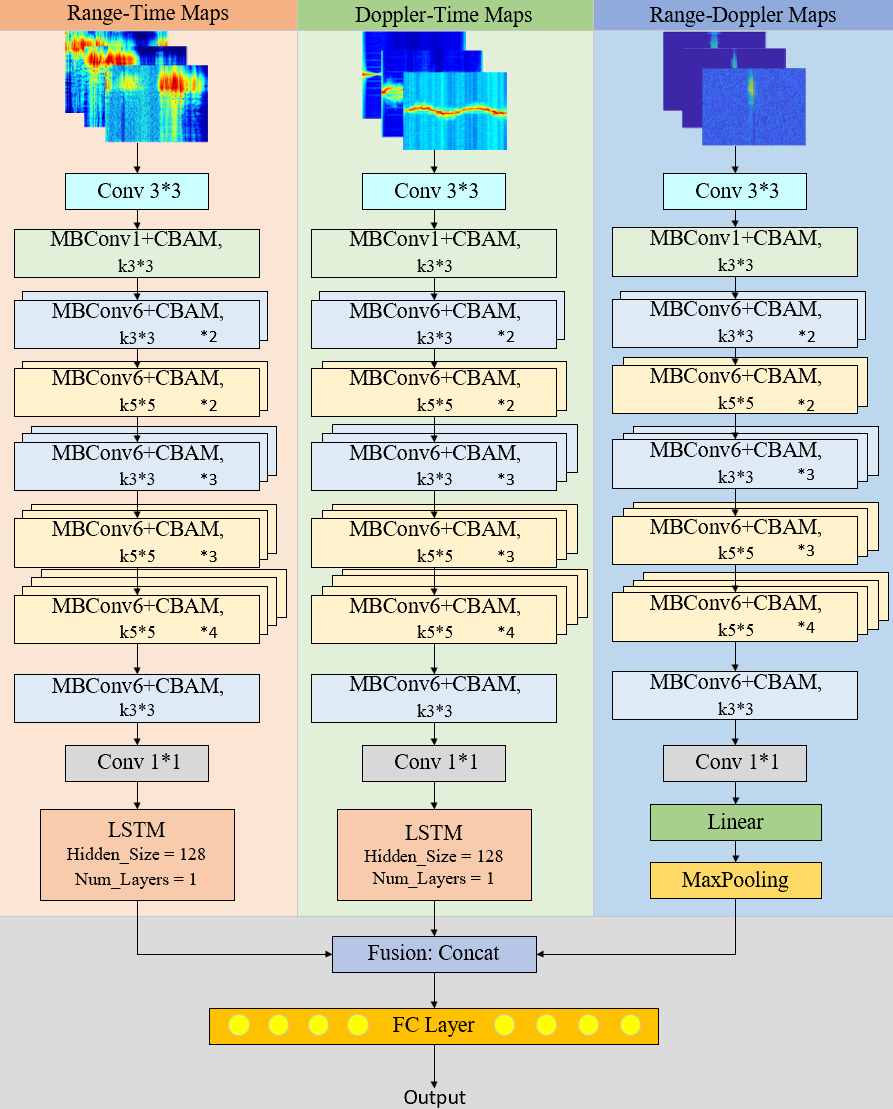}
    \caption{Architecture of the proposed PECL framework with multi-domain processing branches, spatial attention (CBAM) modules, and temporal sequence (LSTM) module. }
    \label{PECL2}
\end{figure}

Let $f_{RT}, f_{DT}, f_{RD} \in \mathbb{R}^{128}$ denote the feature vectors extracted from the Range-Time, Doppler-Time, and Range-Doppler branches, respectively. The fused feature vector is given by:
\begin{equation}
f_{\text{fused}} = \text{Dropout}([f_{RT} \| f_{DT} \| f_{RD}]) \in \mathbb{R}^{384},
\end{equation}
where $\|$ denotes the concatenation operator. This fused vector is subsequently passed to a classification layer. Figure~\ref{PECL2} illustrates the PECL architecture. 

\subsubsection{Attention Mechanism in MBConv Module}
We replace the SE module in the MBConv module with CBAM to better capture both channel and spatial information. While the SE module enhances channel-level features through global average pooling, it fails to consider spatial dependencies in the feature maps — a critical aspect for radar data classification. The CBAM module addresses this limitation through its dual-attention mechanism, which simultaneously processes channel and spatial dimensions. The channel attention module identifies important channel features, while the spatial attention module highlights key spatial regions. 

The channel attention in CBAM employs a shared multi-layer perceptron (MLP) with a reduction ratio $r = 16$, resulting in a bottleneck structure of $C \rightarrow C/16$. This configuration ensures both dimensional compression and recovery, preserving critical feature channels while emphasizing discriminative information.

Given an input feature map $F \in \mathbb{R}^{C \times H \times W}$, the channel attention $M_c \in \mathbb{R}^{C \times 1 \times 1}$ is computed as:
\begin{equation}
M_c = \sigma(\text{MLP}(\text{AvgPool}(F)) + \text{MLP}(\text{MaxPool}(F))),
\end{equation}
where $\sigma$ denotes the sigmoid activation and MLP refers to the two-layer shared perceptron. 

The spatial attention $M_s \in \mathbb{R}^{1 \times H \times W}$ is then computed as: 
\begin{equation}
M_s = \sigma(f^{7\times7}([\text{AvgPool}_c(F); \text{MaxPool}_c(F)])),
\end{equation}
where $f^{7\times7}(\cdot)$ denotes a convolution with a $7 \times 7$ kernel, 
and AvgPool$_c$ and MaxPool$_c$ represent global average pooling and global max pooling along the channel dimension, respectively.

The final attention-enhanced feature is:
\begin{equation}
F' = M_s \cdot (M_c \cdot F).
\end{equation}

\begin{figure}[H]
    \centering
    \includegraphics[scale=0.19]{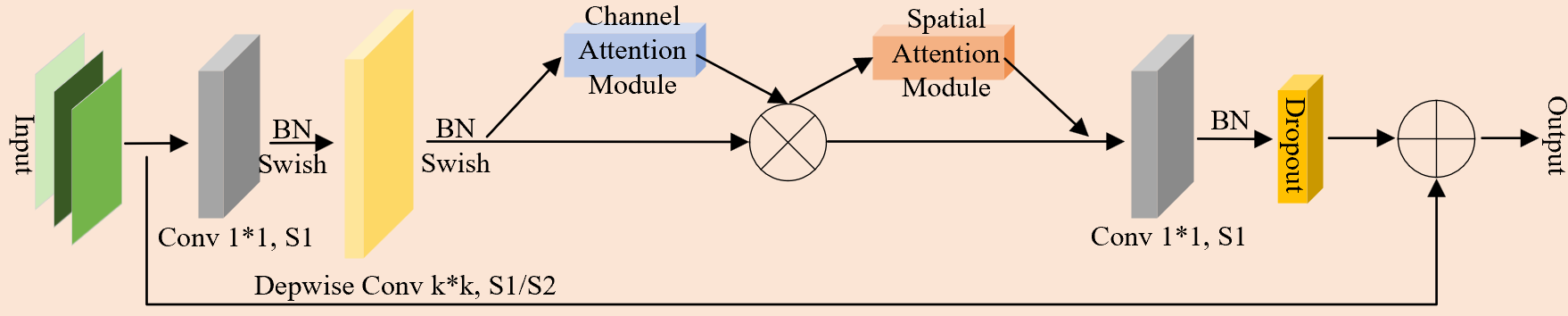}
    \caption{Enhanced MBConv module integrated with CBAM. }
    \label{MBConv}
\end{figure}

This modification brings significant benefits without substantially increasing computational complexity. For instance, when implemented in ResNet50 \cite{woo2018cbam}, both SE and CBAM modules maintain the same parameter count of 28.09M. The computational overhead is nearly identical, with CBAM increasing GFLOPs by only 0.004 (from 3.860 to 3.864). Figure~\ref{MBConv} illustrates the modified MBConv module with CBAM integration. 

\subsubsection{Continuous Temporal Dynamics Extraction}
The 'pick up' and 'drink' actions exhibit strong visual similarity across Range-Time, Doppler-Time, and Range-Doppler maps, resulting in relatively poor classification performance compared to other action categories. To address this challenge, we employ LSTM modules in both the Range-Time and Doppler-Time branches to capture the continuous temporal dynamics of these actions. 

Each branch produces a feature map 
\[
F_d \in \mathbb{R}^{B \times C \times H \times W},
\]
where $B$ is the batch size, $C$ is the number of channels, and $H \times W$ are the spatial dimensions. 
To model temporal evolution, the feature map is reshaped into a sequence: 
\begin{equation}
S_d = \text{Reshape}(F_d) \in \mathbb{R}^{B \times T \times D},
\end{equation}
where $T = W$ is interpreted as the temporal dimension (number of time steps) and $D = H \times C$ is the feature dimension at each step. 
This transformation converts the 2D feature map into a time-ordered sequence suitable for LSTM processing. 

The LSTM processes this sequence and outputs a hidden state $H_d \in \mathbb{R}^{B \times 128}$ corresponding to the last time step: 

\begin{equation}
H_d = \text{LSTM}(S_d)[:, -1, :], \quad d \in \{\text{RT}, \text{DT}\}.
\end{equation}

\begin{figure}[htb]
    \centering
    \includegraphics[scale=0.26]{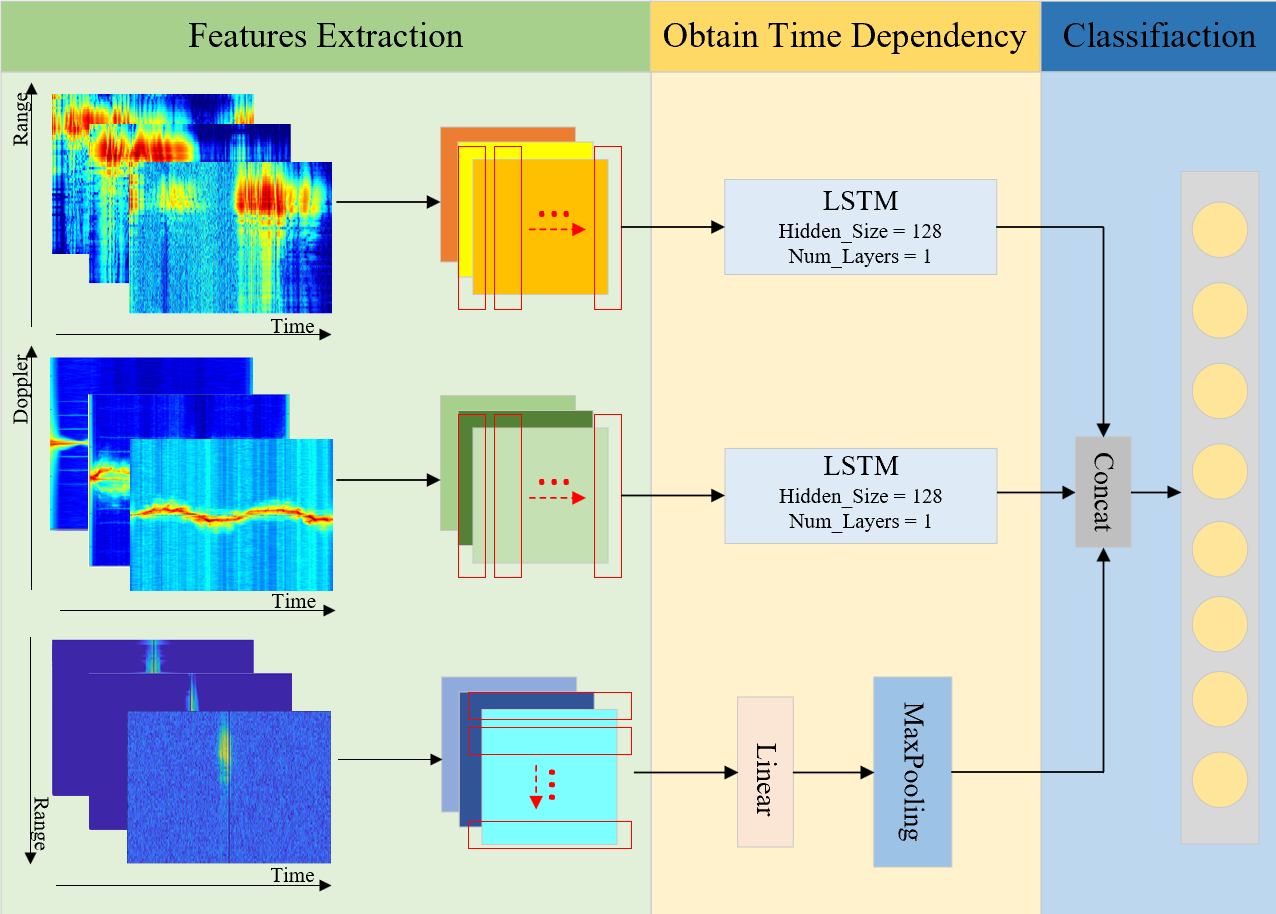}
    \caption{Heterogeneous design for multi-domain feature extraction: Featuring LSTM branches for Range-Time and Doppler-Time maps, and a linear layer for the Range-Doppler maps. }
    \label{LSTM}
\end{figure}

Figure~\ref{LSTM} illustrates the modification of 3 branches by LSTM, depicting the data flow and components of the temporal-domain branches. 

The Range-Doppler domain lacks temporal features, so this branch uses a linear layer followed by max pooling. The linear layer maps features into a compact space, while max pooling highlights the most salient information. This design preserves feature integrity and ensures dimensional consistency with the LSTM outputs from the other branches. 

\section{Implementation and Experimental details}

\subsection{Data Source}

The dataset used in this study is a open source radar dataset from the University of Glasgow \cite{fioranelli2019radar2}, collected with an Ancortek FMCW radar. The system operates at 5.8 GHz (C-band) with 400 MHz bandwidth, 1 ms chirp duration, and +18 dBm output power. Data were recorded in multiple indoor settings with 20 volunteers performing six daily activities: walking, sitting down, standing up, picking up an object, drinking from a cup, and falling. Each subject repeated the actions two to three times, yielding 278, 289, 284, 287, 286, and 198 samples per class, respectively. The raw radar returns were stored as .dat files, each containing a 1D complex array. The first four entries specify system parameters (carrier frequency, chirp duration, number of samples, and bandwidth), and the remaining entries record the echo signals. Experiments were conducted in laboratory rooms, community centers, and residential care facilities. 

\subsection{Data Augmentation}

To mitigate the limited dataset size and the high similarity between certain classes (particularly 'pick up' and 'drink'), we introduce an adaptive noise injection scheme guided by the micro-Doppler energy distribution.

Each spectrogram is divided into three regions according to relative power: low (\textless 30\% of the peak), medium (30–60\%), and high (\textgreater 60\%). Gaussian noise is then injected with different intensities: strong (variance = 1) in low-power regions, moderate in medium regions, and none in high-power regions. This design preserves salient motion features while increasing data diversity.

The strategy is motivated by two observations: (1) radar background noise typically follows a Gaussian distribution, and (2) Gaussian perturbation introduces smooth spectral variation without aliasing. Empirical analysis further showed that low-power bands are more susceptible to noise, justifying stronger perturbation in these regions. 

\subsection{Parameters of PECL and Training Details} 

The PECL network is constructed on EfficientNet-B0 backbones, chosen for their balance between accuracy and efficiency. Each branch begins with a $3\times3$ convolutional stem and progresses through a sequence of MBConv blocks with increasing depth and width (from 16 to 320 channels), followed by a $1\times1$ convolution expanding to 1280 channels. The MBConv blocks are augmented with the Convolutional Block Attention Module (CBAM), where channel attention is implemented via global average and max pooling followed by an MLP with a reduction ratio of $r=16$, and spatial attention is realized through a $7\times7$ convolution over pooled features. 

Temporal modeling differs across domains: the Range-Time and Doppler-Time branches employ a single-layer LSTM with hidden size 128, while the Range-Doppler branch applies a linear projection and max pooling to capture salient spatial cues. Each branch outputs a 128-dimensional feature vector, and the concatenated 384-dimensional representation is regularized with dropout ($p=0.2$) before classification. The detailed layer-by-layer configuration is summarized in Table~\ref{table:pecl-parameters}.

\begin{table*}[htbp]
\centering
\caption{Detailed architecture of the PECL network modules. The backbone is based on EfficientNet-B0.}
\label{table:pecl-parameters}
\renewcommand{\arraystretch}{1.2}
\resizebox{\textwidth}{!}{
\begin{tabular}{l|ll|l}
\hline \textbf{Module} & \textbf{Components} & \textbf{Details} & \textbf{Output Size} \\ \hline
Backbone & Stem Conv & 32 filters, 3$\times$3, stride 2 & (B, 32, 112, 112) \\
         & MBConv1 (Stage 1) & 16 filters, kernel size = 3$\times$3, expand ratio = 1, stride = 1 & (B, 16, 112, 112) \\
         & MBConv6 (Stage 2) & 24 filters, kernel size = 3$\times$3, expand ratio = 6, stride = 2 & (B, 24, 56, 56) \\
         & MBConv6 (Stage 3) & 40 filters, kernel size = 5$\times$5, expand ratio = 6, stride = 2 & (B, 40, 28, 28) \\
         & MBConv6 (Stage 4) & 80 filters, kernel size = 3$\times$3, expand ratio = 6, stride = 2 & (B, 80, 14, 14) \\
         & MBConv6 (Stage 5) & 112 filters, kernel size = 5$\times$5, expand ratio = 6, stride = 1 & (B, 112, 14, 14) \\
         & MBConv6 (Stage 6) & 192 filters, kernel size = 5$\times$5, expand ratio = 6, stride = 2 & (B, 192, 7, 7) \\
         & MBConv6 (Stage 7) & 320 filters, kernel size = 3$\times$3, expand ratio = 6, stride = 1 & (B, 320, 7, 7) \\
         & Conv & 1280 filters, kernel size = 1$\times$1, stride = 1 & (B, 1280, 7, 7) \\ \hline
CBAM     & Channel Attention & Avg+Max Pool, MLP: $C \rightarrow C/16 \rightarrow C$, ReLU & (B, C, 1, 1) \\
         & Spatial Attention & 7$\times$7 Conv on [Avg, Max] & (B, 1, H, W) \\ \hline
Temporal Module & LSTM (RT/DT) & Input: (B, 7, 1280), 1 layer, hidden size 128 & (B, 128) \\
                & Linear+MaxPool (RD) & Linear(8960→128), max over time dim & (B, 128) \\ \hline
Fusion   & Concatenate & [RT, DT, RD] $\in \mathbb{R}^{128 \times 3}$ & (B, 384) \\
         & FC + Dropout & Dropout(0.2), Linear(384→6) & (B, 6) \\ \hline
\end{tabular}}
\end{table*}

The models were trained and evaluated on an NVIDIA A100 GPU. Data were split into training (60\%), validation (20\%), and test (20\%) subsets. Optimization was performed using the Adam optimizer with an initial learning rate of 0.001, decayed by a factor of 0.1 every 30 epochs. Cross-entropy loss was employed as the objective function, and training proceeded for 300 epochs with a mini-batch size of 32. Batch normalization was applied after convolutional layers to stabilize training, and a dropout rate of 0.2 was used in the fusion module to reduce overfitting. 

\section{Results and Ablation Experiment}
\subsection{Result Analysis}

The proposed PECL model achieved an overall accuracy of 96.16\% on our dataset. Specifically, PECL reached accuracies of 100.00\%, 98.94\%, 100.00\%, and 98.61\% for the categories \textit{Walking}, \textit{Sitting Down}, \textit{Stand Up}, and \textit{Fall}, respectively. The actions \textit{Pick Up} and \textit{Drink} were relatively more challenging to distinguish, yet the model still achieved high accuracies of 89.77\% and 90.38\% for these classes.

PECL contains 23.42 million parameters and requires 1324.82 million FLOPs, indicating that it remains computationally efficient while maintaining strong recognition performance. In contrast, most ViT-based architectures trade off model compactness for improved accuracy, resulting in substantially higher computational costs. By leveraging a parallel multi-branch design, PECL enhances feature extraction capacity while preventing excessive model complexity.

As shown in Table~\ref{table:comparison_all}, PECL achieves a strong balance between classification performance and model efficiency. The brightness of colors in the chart provides a visual representation of accuracy levels. Brighter colors indicate higher accuracy. Overall, PECL outperforms both classical CNNs and recent transformer-based approaches in terms of both recognition precision and computational efficiency.

\begin{table*}[htbp]
\centering
\caption{Performance (\%) and complexity comparison: PECL versus classical and contemporary models. }
\resizebox{\textwidth}{!}{
\renewcommand{\arraystretch}{1.15}
\setlength{\tabcolsep}{4pt}
\begin{tabular}{c|cccccc|c|cc}
\hline
\textbf{Model} & \textbf{Walk} & \textbf{Sit} & \textbf{Stand} & \textbf{Pick} & \textbf{Drink} & \textbf{Fall} & \textbf{Acc.} & \textbf{Params (M)} & \textbf{FLOPs (M)} \\ 
\hline
LightWeightNet \cite{ou2022lightweight} & \cellcolor{gray!70}69.39 & \cellcolor{gray!20}97.20 & \cellcolor{gray!60}71.07 & \cellcolor{gray!65}62.50 & \cellcolor{gray!75}57.55 & \cellcolor{gray!40}85.23 & \cellcolor{gray!65}72.31 & 0.51 & 75.06 \\
2DCNN+CBAM \cite{he2023fall} & \cellcolor{gray!25}96.43 & \cellcolor{gray!35}90.67 & \cellcolor{gray!30}91.38 & \cellcolor{gray!55}73.33 & \cellcolor{gray!70}58.21 & \cellcolor{gray!10}100.00 & \cellcolor{gray!40}84.62 & 38.61 & 453.18 \\
Custom 14-Layer ResNet \cite{janakaraj2019star} & \cellcolor{gray!20}98.00 & \cellcolor{gray!30}91.80 & \cellcolor{gray!35}89.29 & \cellcolor{gray!45}82.61 & \cellcolor{gray!55}73.68 & \cellcolor{gray!10}100.00 & \cellcolor{gray!35}87.69 & 13.67 & 1939.36 \\
ResNext50\_32x4D & \cellcolor{gray!25}96.36 & \cellcolor{gray!10}100.00 & \cellcolor{gray!30}92.16 & \cellcolor{gray!55}75.00 & \cellcolor{gray!55}73.77 & \cellcolor{gray!10}100.00 & \cellcolor{gray!30}88.31 & 25.03 & 4288.18 \\
ShuffleNet & \cellcolor{gray!10}100.00 & \cellcolor{gray!30}91.80 & \cellcolor{gray!15}98.11 & \cellcolor{gray!50}75.93 & \cellcolor{gray!50}76.06 & \cellcolor{gray!10}100.00 & \cellcolor{gray!25}88.92 & 2.28 & 152.71 \\
VGG16 & \cellcolor{gray!35}92.65 & \cellcolor{gray!10}100.00 & \cellcolor{gray!25}94.44 & \cellcolor{gray!45}80.00 & \cellcolor{gray!35}82.76 & \cellcolor{gray!20}97.44 & \cellcolor{gray!25}90.77 & 138.36 & 15470.31 \\
MobileNet\_V3\_Large & \cellcolor{gray!30}95.38 & \cellcolor{gray!10}100.00 & \cellcolor{gray!25}93.44 & \cellcolor{gray!35}89.29 & \cellcolor{gray!55}74.24 & \cellcolor{gray!10}100.00 & \cellcolor{gray!25}90.77 & 5.48 & 234.84 \\
MobileNet\_V2 & \cellcolor{gray!20}98.04 & \cellcolor{gray!25}96.00 & \cellcolor{gray!15}98.46 & \cellcolor{gray!40}82.81 & \cellcolor{gray!45}77.36 & \cellcolor{gray!20}97.62 & \cellcolor{gray!20}91.38 & 3.50 & 327.49 \\
ResNet50 & \cellcolor{gray!10}100.00 & \cellcolor{gray!20}98.28 & \cellcolor{gray!15}98.36 & \cellcolor{gray!40}85.96 & \cellcolor{gray!55}74.65 & \cellcolor{gray!10}100.00 & \cellcolor{gray!20}91.38 & 25.56 & 4133.74 \\ 
\hline
ViT\_B\_16 & \cellcolor{gray!10}100.00 & \cellcolor{gray!25}96.36 & \cellcolor{gray!30}95.71 & \cellcolor{gray!45}80.95 & \cellcolor{gray!60}69.23 & \cellcolor{gray!35}94.59 & \cellcolor{gray!30}89.23 & 58.07 & 11286.25 \\
ViT\_B\_32 & \cellcolor{gray!25}96.83 & \cellcolor{gray!35}90.91 & \cellcolor{gray!25}97.96 & \cellcolor{gray!40}85.71 & \cellcolor{gray!45}78.00 & \cellcolor{gray!10}100.00 & \cellcolor{gray!20}91.08 & 59.84 & 2951.37 \\ 
\hline
\rowcolor{gray!15}
\textbf{PECL (Ours)} & 100.00 & 98.94 & 100.00 & 89.77 & 90.38 & 98.61 & 96.16 & 23.42 & 1324.82 \\ 
\hline
\end{tabular}}
\label{table:comparison_all}
\end{table*}

\subsection{Verify the Generalization Ability of Model}

The generalization capability of the proposed approach was further evaluated using an external open dataset~\cite{gurbuz2020cross}, which contains only Doppler–Time (D–T) spectrograms across 11 activity classes (60 samples per class). The complete Parallel-EfficientNet-CBAM-LSTM (PECL) architecture comprises three parallel EfficientNet-B0 (Eff\_B0) branches processing Range–Time (R–T), Doppler–Time (D–T), and Range–Doppler (R–D) representations, where each MBConv block replaces its SE module with a CBAM and LSTM units are attached to the R–T and D–T branches. Because the external dataset provides only D–T inputs, the evaluation focused on the D–T branch of PECL, denoted as “Eff\_B0+CBAM+LSTM.” This variant integrates an EfficientNet-B0 backbone with CBAM attention and the same temporal LSTM module used in the full model, where the D–T maps is segmented along the time axis and the final hidden state is used for classification.

For a fair comparison, this single-branch model was trained under the same optimization settings as described in Section~IV and evaluated against representative CNN architecture of comparable scale. Networks based on the Transformer architecture perform poorly on this small dataset. As summarized in Table~\ref{table:Generalization}, the proposed D–T branch achieved the highest overall accuracy of 84.85\%, outperforming all competing baselines. The lowest per-class performance was observed in the “Towards” category (68.18\%), likely due to strong inter-class similarity in Doppler–Time signatures and the limited sample count per class. Other classes exhibit lighter and darker shades in the table, reflecting stable accuracy across most actions and confirming the consistent generalization.

The superior performance of Eff\_B0+CBAM+LSTM arises from three main design choices: (1) the EfficientNet-B0 backbone achieves an optimal trade-off between representational capacity and parameter efficiency on small datasets; (2) CBAM effectively enhances salient spectral–spatial features while suppressing background noise; (3) LSTM captures temporal evolution along the Doppler–Time axis, a property absent in pure CNN baselines. 

Although the complete multi-domain PECL architecture cannot be directly evaluated on this dataset, results in Table~\ref{table:Parallel_Structure} demonstrate that combining R–T, D–T, and R–D branches further enhances accuracy. Consequently, the strong performance of the D–T branch alone supports two conclusions: (i) each PECL sub-branch functions as a capable and transferable feature extractor, and (ii) multi-domain fusion is expected to yield even stronger cross-dataset generalization when complete domain information is available. 

\begin{table*}[htbp]
\centering
\caption{Comparison of generalization ability across multiple activity classes (\%).}
\resizebox{\textwidth}{!}{
\renewcommand{\arraystretch}{1.15}
\setlength{\tabcolsep}{4pt}
\begin{tabular}{c|ccccccccccc|c}
\hline
\textbf{Model} & \textbf{Away} & \textbf{Bend} & \textbf{Crawl} & \textbf{Kneel} & \textbf{Limp} & \textbf{Pick} & \textbf{Step} & \textbf{Scissor} & \textbf{Sit} & \textbf{Toes} & \textbf{Towards} & \textbf{Acc.} \\ 
\hline
MobileNet\_V2 & \cellcolor{gray!10}100.00 & \cellcolor{gray!55}71.43 & \cellcolor{gray!10}100.00 & \cellcolor{gray!35}80.00 & \cellcolor{gray!70}50.00 & \cellcolor{gray!60}73.33 & \cellcolor{gray!25}93.33 & \cellcolor{gray!80}35.71 & \cellcolor{gray!65}53.85 & \cellcolor{gray!70}50.00 & \cellcolor{gray!65}54.17 & \cellcolor{gray!55}68.15 \\
ResNet50 & \cellcolor{gray!10}100.00 & \cellcolor{gray!10}100.00 & \cellcolor{gray!10}100.00 & \cellcolor{gray!10}100.00 & \cellcolor{gray!25}85.71 & \cellcolor{gray!60}66.67 & \cellcolor{gray!70}48.65 & \cellcolor{gray!10}100.00 & \cellcolor{gray!25}90.91 & \cellcolor{gray!70}44.00 & \cellcolor{gray!40}76.92 & \cellcolor{gray!50}71.97 \\
ResNext50\_32x4D & \cellcolor{gray!10}100.00 & \cellcolor{gray!10}100.00 & \cellcolor{gray!25}92.86 & \cellcolor{gray!45}68.00 & \cellcolor{gray!85}25.00 & \cellcolor{gray!60}72.73 & \cellcolor{gray!10}100.00 & \cellcolor{gray!65}50.00 & \cellcolor{gray!10}100.00 & \cellcolor{gray!10}100.00 & \cellcolor{gray!60}62.50 & \cellcolor{gray!45}73.89 \\ 
ShuffleNet & \cellcolor{gray!25}90.00 & \cellcolor{gray!45}80.00 & \cellcolor{gray!25}93.75 & \cellcolor{gray!35}85.71 & \cellcolor{gray!50}63.64 & \cellcolor{gray!45}80.00 & \cellcolor{gray!25}94.44 & \cellcolor{gray!65}50.00 & \cellcolor{gray!10}100.00 & \cellcolor{gray!70}39.13 & \cellcolor{gray!35}83.33 & \cellcolor{gray!40}75.80 \\
Custom 14-Layer ResNet & \cellcolor{gray!10}100.00 & \cellcolor{gray!35}88.89 & \cellcolor{gray!10}100.00 & \cellcolor{gray!10}100.00 & \cellcolor{gray!25}85.71 & \cellcolor{gray!40}85.00 & \cellcolor{gray!45}70.59 & \cellcolor{gray!10}100.00 & \cellcolor{gray!25}87.50 & \cellcolor{gray!55}66.67 & \cellcolor{gray!70}44.12 & \cellcolor{gray!35}77.07 \\
LightWeightNet & \cellcolor{gray!55}72.41 & \cellcolor{gray!60}68.75 & \cellcolor{gray!55}73.33 & \cellcolor{gray!65}54.55 & \cellcolor{gray!35}86.49 & \cellcolor{gray!60}70.59 & \cellcolor{gray!25}92.31 & \cellcolor{gray!10}100.00 & \cellcolor{gray!15}96.15 & \cellcolor{gray!15}96.77 & \cellcolor{gray!45}69.23 & \cellcolor{gray!35}77.07 \\ 
\hline
\rowcolor{gray!15}
\textbf{Eff\_B0+CBAM+LSTM (Ours)} & 100.00 & 84.21 & 72.73 & 91.30 & 75.00 & 94.44 & 84.71 & 92.86 & 94.12 & 80.00 & 68.18 & 84.85 \\ 
\hline
\end{tabular}}
\label{table:Generalization}
\end{table*}

\subsection{Ablation Study}
A series of ablation experiments were conducted to examine the contribution of the parallel architecture, CBAM attention, and LSTM temporal modeling to classification performance. We employ heatmap-style tables and confusion matrices to demonstrate model accuracy, and visualize classification performance using t-SNE. 

\subsubsection{Advantages of Parallel Architecture}
To evaluate the effect of multi-domain fusion, we compared single-domain inputs against the proposed parallel structure. As presented in Table~\ref{table:Parallel_Structure}, the Parallel-EfficientNet achieved an overall accuracy of 91.08\%, representing improvements of 1.54\%, 17.54\%, and 7.39\% over single-domain EfficientNet-B0 models trained on R–T, D–T, and R–D inputs, respectively.

Among all categories, “Pick Up” benefits the most, with its accuracy improving by up to 27.13\% relative to the weakest single-domain baseline. This demonstrates that simultaneous learning from R–T, D–T, and R–D domains provides richer spatial–temporal cues for distinguishing actions with subtle temporal overlaps. 

Figures.~\ref{fig:tsne_all}(b)–(e) further illustrate the embedding distributions using t-SNE visualization. While single-domain models show partial overlap—especially between “Stand Up,” “Pick Up,” and “Drink”—the feature clusters in the parallel architecture (Figure~\ref{fig:tsne_all}(e)) become more distinct, confirming that multi-domain fusion yields a more discriminative latent space. 

\begin{table}[h]
\centering
\caption{Ablation Study on the multi-branch parallel design (\%). }
\resizebox{\columnwidth}{!}{
\renewcommand{\arraystretch}{1.15}
\setlength{\tabcolsep}{3pt}
\begin{tabular}{c|cccccc|c}
\hline
\textbf{Model} & \textbf{Walking} & \textbf{Sitting} & \textbf{Stand up} & \textbf{Pick up} & \textbf{Drink} & \textbf{Fall} & \textbf{Acc.} \\
\hline
Eff\_B0 (R--T) & \cellcolor{gray!15}97.78 & \cellcolor{gray!70}68.00 & \cellcolor{gray!60}71.43 & \cellcolor{gray!75}58.51 & \cellcolor{gray!55}73.47 & \cellcolor{gray!20}96.77 & \cellcolor{gray!55}73.54 \\
Eff\_B0 (D--T) & \cellcolor{gray!10}100.00 & \cellcolor{gray!20}94.55 & \cellcolor{gray!25}95.59 & \cellcolor{gray!45}79.07 & \cellcolor{gray!65}67.80 & \cellcolor{gray!10}100.00 & \cellcolor{gray!25}89.54 \\
Eff\_B0 (R--D) & \cellcolor{gray!25}96.15 & \cellcolor{gray!20}95.56 & \cellcolor{gray!35}85.71 & \cellcolor{gray!65}61.11 & \cellcolor{gray!55}74.65 & \cellcolor{gray!10}100.00 & \cellcolor{gray!40}83.69 \\
\hline
\textbf{Parallel-Eff\_B0} & \cellcolor{gray!20}95.65 & \cellcolor{gray!30}92.31 & \cellcolor{gray!15}98.28 & \cellcolor{gray!35}88.24 & \cellcolor{gray!50}75.44 & \cellcolor{gray!25}97.37 & \cellcolor{gray!20}\textbf{91.08} \\
\hline
\end{tabular}}
\label{table:Parallel_Structure}
\end{table}

\subsubsection{Replacement of SE Module with CBAM}

We next examined the effect of replacing the Squeeze-and-Excitation (SE) module with the Convolutional Block Attention Module (CBAM) in the parallel EfficientNet framework. As shown in Table~\ref{table:CBAM}, the Parallel-EfficientNet-CBAM (PEC) consistently outperformed its SE-based counterpart across all branches.

For single-domain branches, accuracies improved from 73.54\% to 74.46\% in the Range–Time (R–T) domain, from 89.54\% to 90.15\% in the Doppler–Time (D–T) domain, and from 83.69\% to 85.23\% in the Range–Doppler (R–D) domain. The complete parallel network benefited most, achieving an overall accuracy increase from 91.08\% to 92.62\%. These consistent gains confirm that CBAM’s joint channel–spatial attention captures more discriminative features than the SE module’s channel-only mechanism—particularly for activities that exhibit subtle spatial and temporal cues. 

We use t-SNE to visualize further support this finding. Figure~\ref{fig:tsne_all}(f) illustrates that CBAM leads to clearer separation between action classes, especially in challenging categories such as “Pick up” and “Drink,” which remain closely clustered under SE.

\begin{table}[h]
\centering
\caption{Ablation study on integrating CBAM in multi-domain branches (\%). }
\resizebox{\columnwidth}{!}{
\renewcommand{\arraystretch}{1.15}
\setlength{\tabcolsep}{4pt}
\begin{tabular}{c|cccccc|c}
\hline
\textbf{Model} & \textbf{Walk} & \textbf{Sit} & \textbf{Stand} & \textbf{Pick} & \textbf{Drink} & \textbf{Fall} & \textbf{Acc.} \\
\hline
Eff\_B0 (R–T) & \cellcolor{gray!20}97.78 & \cellcolor{gray!65}68.00 & \cellcolor{gray!60}71.43 & \cellcolor{gray!70}58.51 & \cellcolor{gray!55}73.47 & \cellcolor{gray!25}96.77 & \cellcolor{gray!60}73.54 \\
Eff\_B0 (D–T) & \cellcolor{gray!10}100.00 & \cellcolor{gray!35}94.55 & \cellcolor{gray!30}95.59 & \cellcolor{gray!50}79.07 & \cellcolor{gray!55}67.80 & \cellcolor{gray!10}100.00 & \cellcolor{gray!35}89.54 \\
Eff\_B0 (R–D) & \cellcolor{gray!30}96.15 & \cellcolor{gray!35}95.56 & \cellcolor{gray!45}85.71 & \cellcolor{gray!60}61.11 & \cellcolor{gray!50}74.65 & \cellcolor{gray!10}100.00 & \cellcolor{gray!45}83.69 \\ \hline
Eff+CBAM (R–T) & \cellcolor{gray!35}93.85 & \cellcolor{gray!40}90.63 & \cellcolor{gray!70}58.46 & \cellcolor{gray!65}62.28 & \cellcolor{gray!60}66.07 & \cellcolor{gray!45}85.71 & \cellcolor{gray!60}74.46 \\
Eff+CBAM (D–T) & \cellcolor{gray!10}100.00 & \cellcolor{gray!20}98.00 & \cellcolor{gray!35}92.98 & \cellcolor{gray!40}85.25 & \cellcolor{gray!45}74.65 & \cellcolor{gray!10}100.00 & \cellcolor{gray!35}90.15 \\
Eff+CBAM (R–D) & \cellcolor{gray!30}94.00 & \cellcolor{gray!40}90.63 & \cellcolor{gray!35}90.16 & \cellcolor{gray!50}74.14 & \cellcolor{gray!55}67.92 & \cellcolor{gray!25}97.44 & \cellcolor{gray!45}85.23 \\ \hline
Parallel-Eff\_B0 & \cellcolor{gray!35}95.65 & \cellcolor{gray!40}92.31 & \cellcolor{gray!20}98.28 & \cellcolor{gray!40}88.24 & \cellcolor{gray!45}75.44 & \cellcolor{gray!25}97.37 & \cellcolor{gray!30}91.08 \\
\textbf{Parallel-Eff}\_B0+CBAM & \cellcolor{gray!10}100.00 & \cellcolor{gray!20}98.46 & \cellcolor{gray!30}94.34 & \cellcolor{gray!40}87.50 & \cellcolor{gray!45}79.71 & \cellcolor{gray!10}100.00 & \cellcolor{gray!25}92.62 \\ 
\hline
\end{tabular}}
\label{table:CBAM}
\end{table}

\subsubsection{Analysis of the impact on the LSTM module and modifications}

To enhance the model's capability in distinguishing between the frequently confused 'pick up' and 'drink' actions, we implemented LSTM modules to capture temporal dependencies in action sequences. As evidenced by the ablation study results presented in Table~\ref{table:LSTM}, the LSTM modules demonstrate domain-specific effectiveness. The Doppler-Time domain branch shows an accuracy improvement from 89.54\% to 90.46\%, while the Range-Time branch achieves a more substantial gain from 73.54\% to 76.00\%. However, the Range-Doppler domain exhibits contrasting behavior, with accuracy decreasing from 83.69\% to 74.46\% when processed through LSTM modules. We incorporated LSTM modules into all three branches. This modification resulted in a decrease in the parallel model's accuracy from 92.62\% to 82.15\%. 

\begin{table}[H]
\centering
\caption{Ablation study on the role of the temporal module (\%). }
\resizebox{\columnwidth}{!}{
\renewcommand{\arraystretch}{1.15}
\setlength{\tabcolsep}{4pt}
\begin{tabular}{c|cccccc|c}
\hline
\textbf{Model} & \textbf{Walk} & \textbf{Sit} & \textbf{Stand} & \textbf{Pick} & \textbf{Drink} & \textbf{Fall} & \textbf{Acc.} \\
\hline
Eff\_B0 (R–T) & \cellcolor{gray!20}97.78 & \cellcolor{gray!65}68.00 & \cellcolor{gray!60}71.43 & \cellcolor{gray!70}58.51 & \cellcolor{gray!55}73.47 & \cellcolor{gray!25}96.77 & \cellcolor{gray!60}73.54 \\
Eff\_B0 (D–T) & \cellcolor{gray!10}100.00 & \cellcolor{gray!35}94.55 & \cellcolor{gray!30}95.59 & \cellcolor{gray!50}79.07 & \cellcolor{gray!55}67.80 & \cellcolor{gray!10}100.00 & \cellcolor{gray!35}89.54 \\
Eff\_B0 (R–D) & \cellcolor{gray!30}96.15 & \cellcolor{gray!35}95.56 & \cellcolor{gray!45}85.71 & \cellcolor{gray!60}61.11 & \cellcolor{gray!50}74.65 & \cellcolor{gray!10}100.00 & \cellcolor{gray!45}83.69 \\ \hline
Eff+LSTM (R–T) & \cellcolor{gray!10}100.00 & \cellcolor{gray!60}70.91 & \cellcolor{gray!55}74.58 & \cellcolor{gray!65}64.00 & \cellcolor{gray!65}62.50 & \cellcolor{gray!30}93.33 & \cellcolor{gray!55}76.00 \\
Eff+LSTM (D–T) & \cellcolor{gray!10}100.00 & \cellcolor{gray!30}96.61 & \cellcolor{gray!25}98.04 & \cellcolor{gray!45}83.67 & \cellcolor{gray!55}71.01 & \cellcolor{gray!10}100.00 & \cellcolor{gray!35}90.46 \\
Eff+LSTM (R–D) & \cellcolor{gray!40}88.52 & \cellcolor{gray!25}97.63 & \cellcolor{gray!35}89.09 & \cellcolor{gray!65}58.33 & \cellcolor{gray!80}50.00 & \cellcolor{gray!40}89.74 & \cellcolor{gray!60}74.46 \\ \hline
PEC & \cellcolor{gray!10}100.00 & \cellcolor{gray!20}98.46 & \cellcolor{gray!30}94.34 & \cellcolor{gray!40}87.50 & \cellcolor{gray!45}79.71 & \cellcolor{gray!10}100.00 & \cellcolor{gray!25}92.62 \\
PEC+3 LSTM & \cellcolor{gray!35}96.00 & \cellcolor{gray!45}82.14 & \cellcolor{gray!55}75.76 & \cellcolor{gray!60}69.23 & \cellcolor{gray!45}80.39 & \cellcolor{gray!10}100.00 & \cellcolor{gray!45}82.15 \\
\rowcolor{gray!15}
\textbf{PECL} & 100.00 & 98.94 & 100.00 & 89.77 & 90.38 & 98.61 & 96.16 \\ 
\hline
\end{tabular}}
\label{table:LSTM}
\end{table}

\begin{figure*}[h]
    \centering
    \begin{minipage}{0.245\linewidth}
        \centerline{\includegraphics[width=\textwidth]{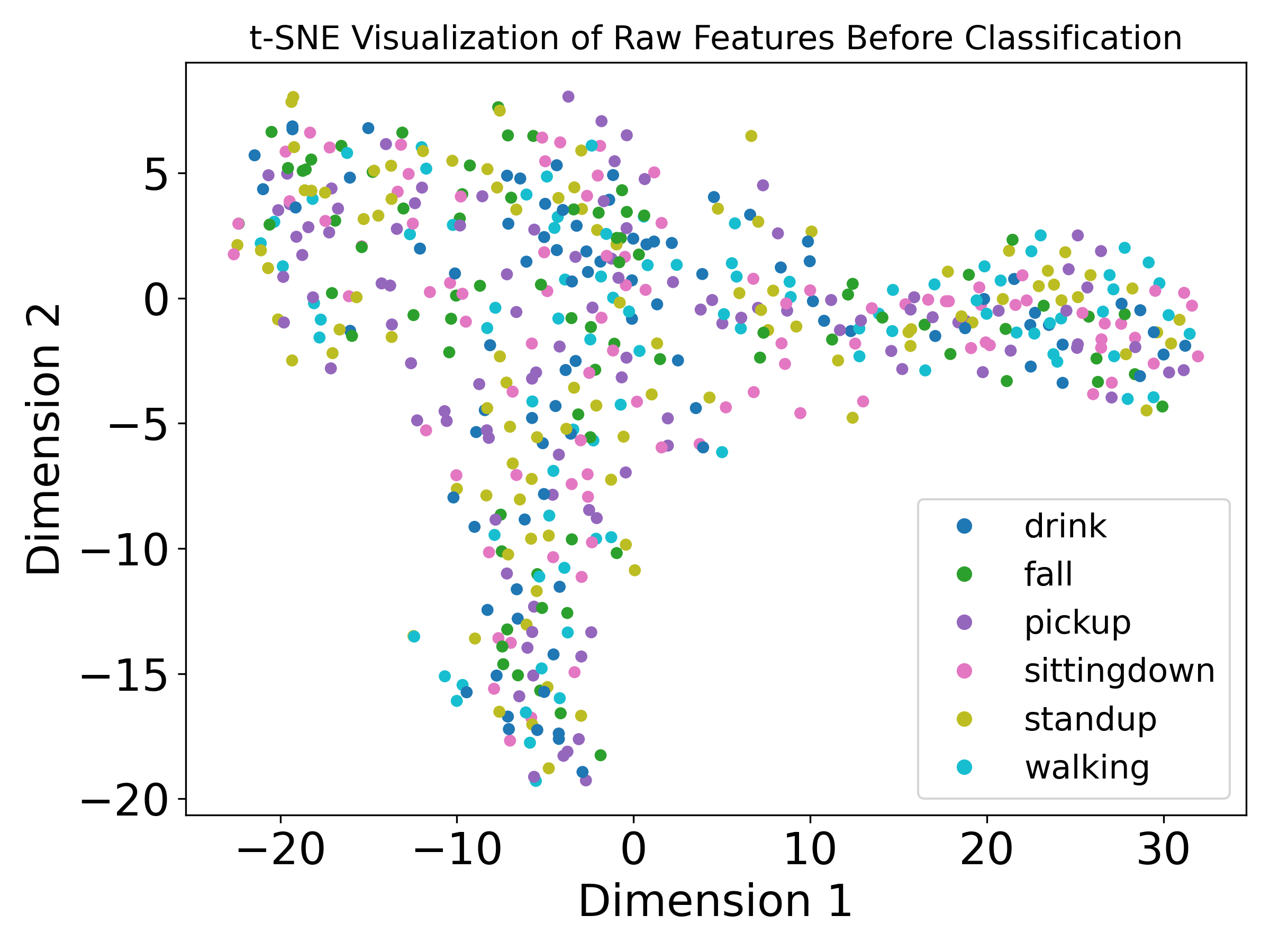}}
        \centerline{(a) \label{fig:tsne_raw}}
    \end{minipage}
    \hfill
    \begin{minipage}{0.245\linewidth}
        \centerline{\includegraphics[width=\textwidth]{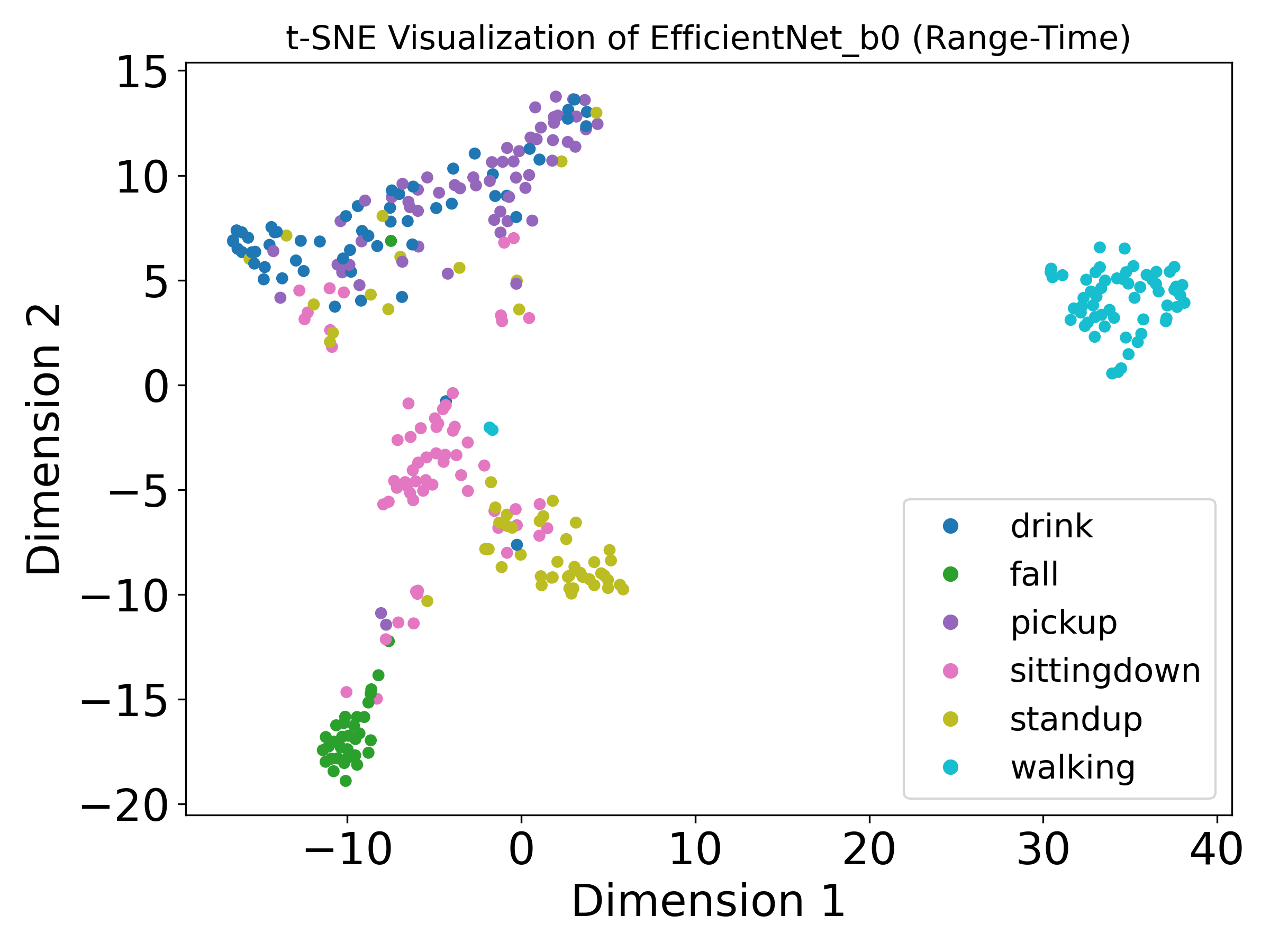}}
        \centerline{(b) \label{fig:tsne_rt}}
    \end{minipage}
    \hfill
    \begin{minipage}{0.245\linewidth}
        \centerline{\includegraphics[width=\textwidth]{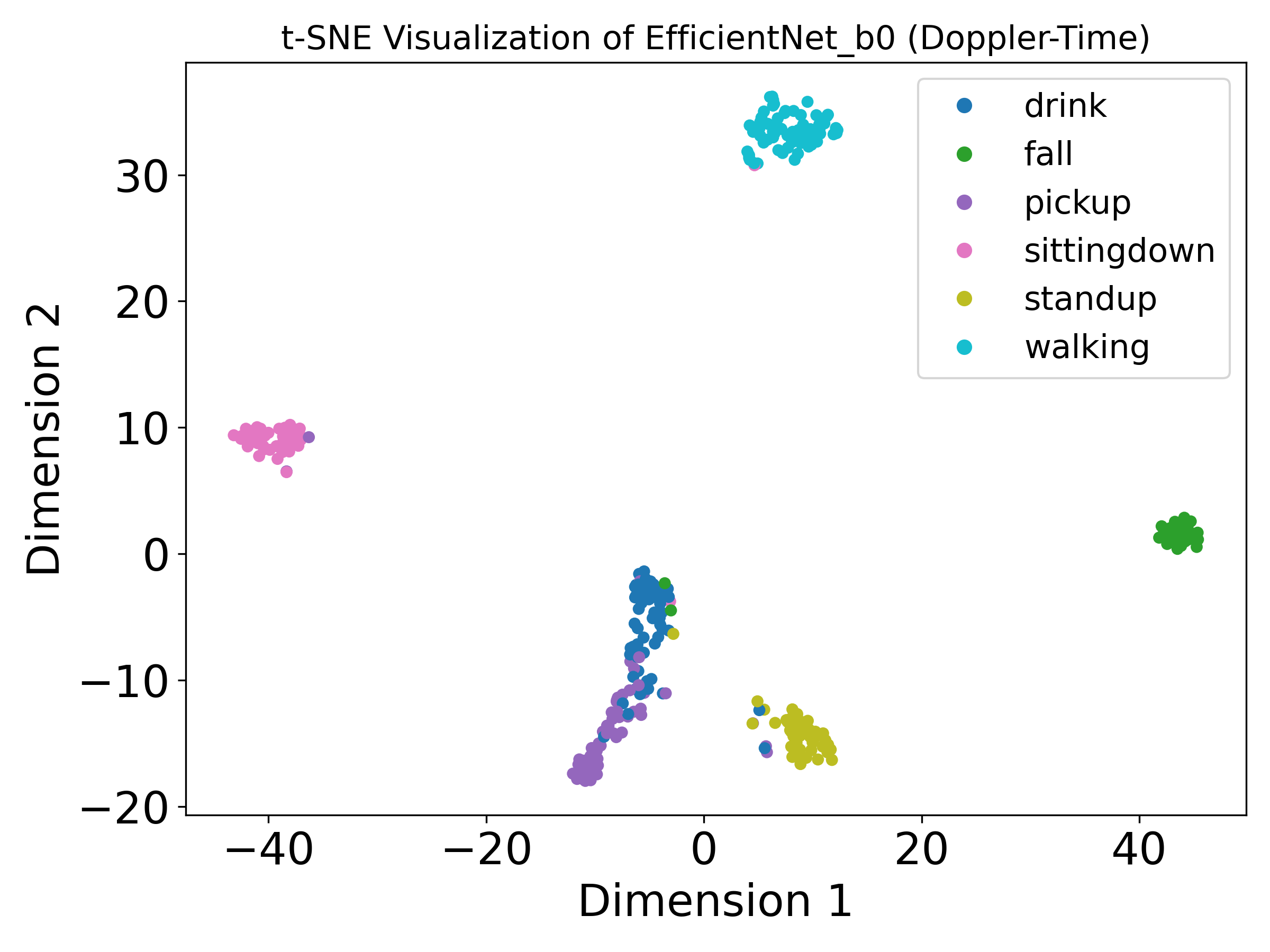}}
        \centerline{(c) \label{fig:tsne_dt}}
    \end{minipage}
    \hfill
    \begin{minipage}{0.245\linewidth}
        \centerline{\includegraphics[width=\textwidth]{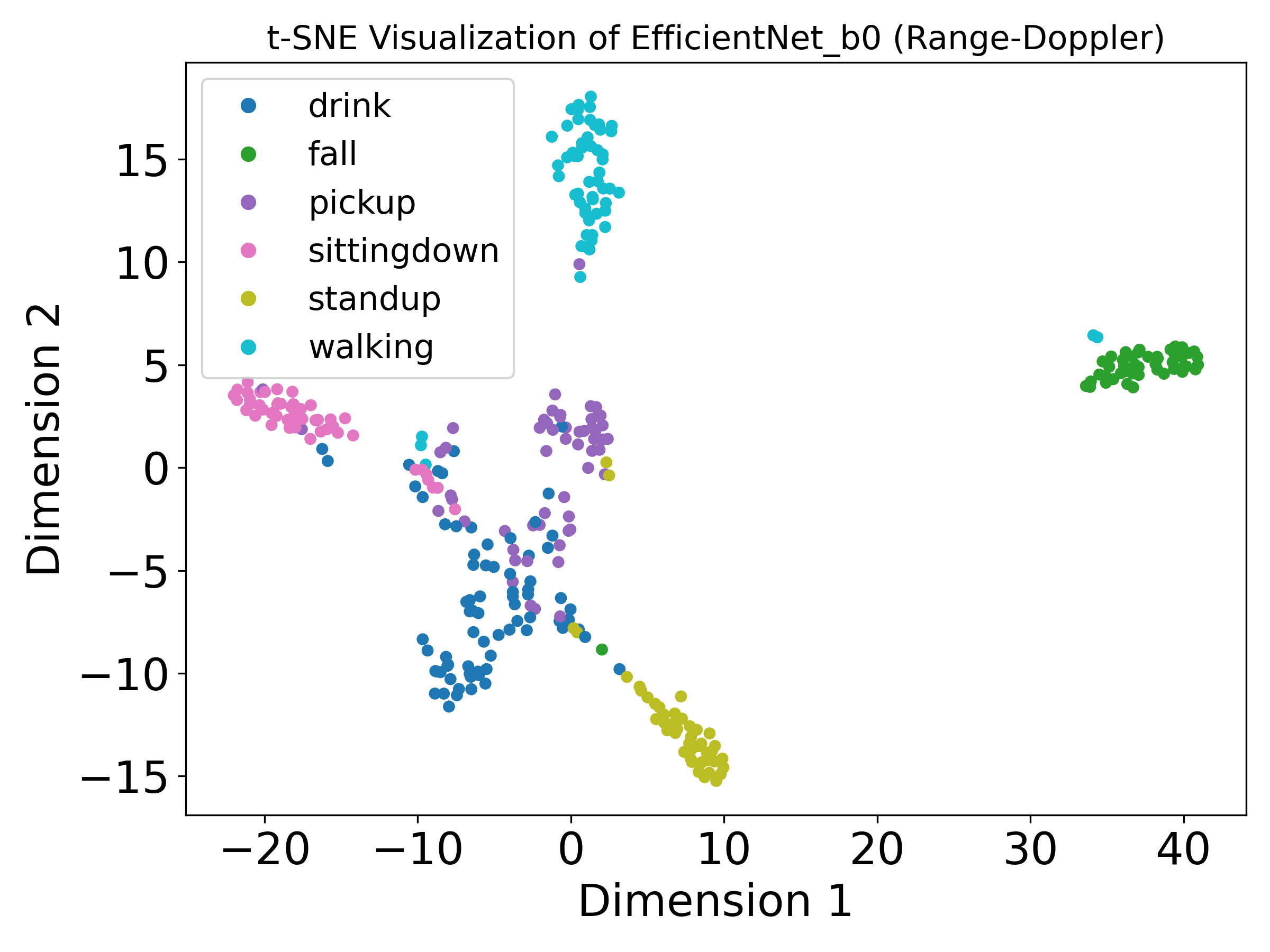}}
        \centerline{(d) \label{fig:tsne_rd}}
    \end{minipage}

    \vspace{4pt}

    \begin{minipage}{0.245\linewidth}
        \centerline{\includegraphics[width=\textwidth]{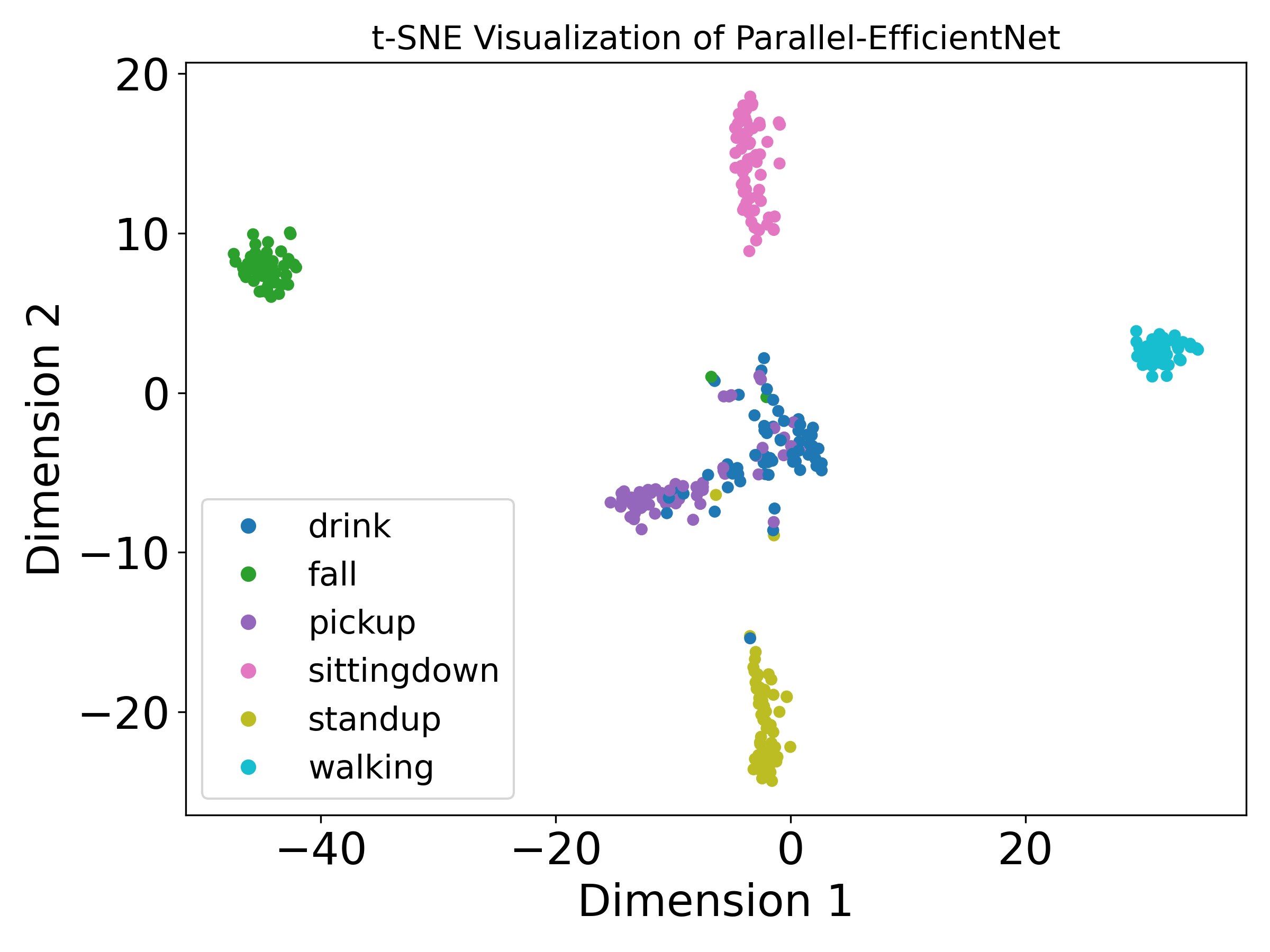}}
        \centerline{(e) \label{fig:tsne_pe}}
    \end{minipage}
    \hfill
    \begin{minipage}{0.245\linewidth}
        \centerline{\includegraphics[width=\textwidth]{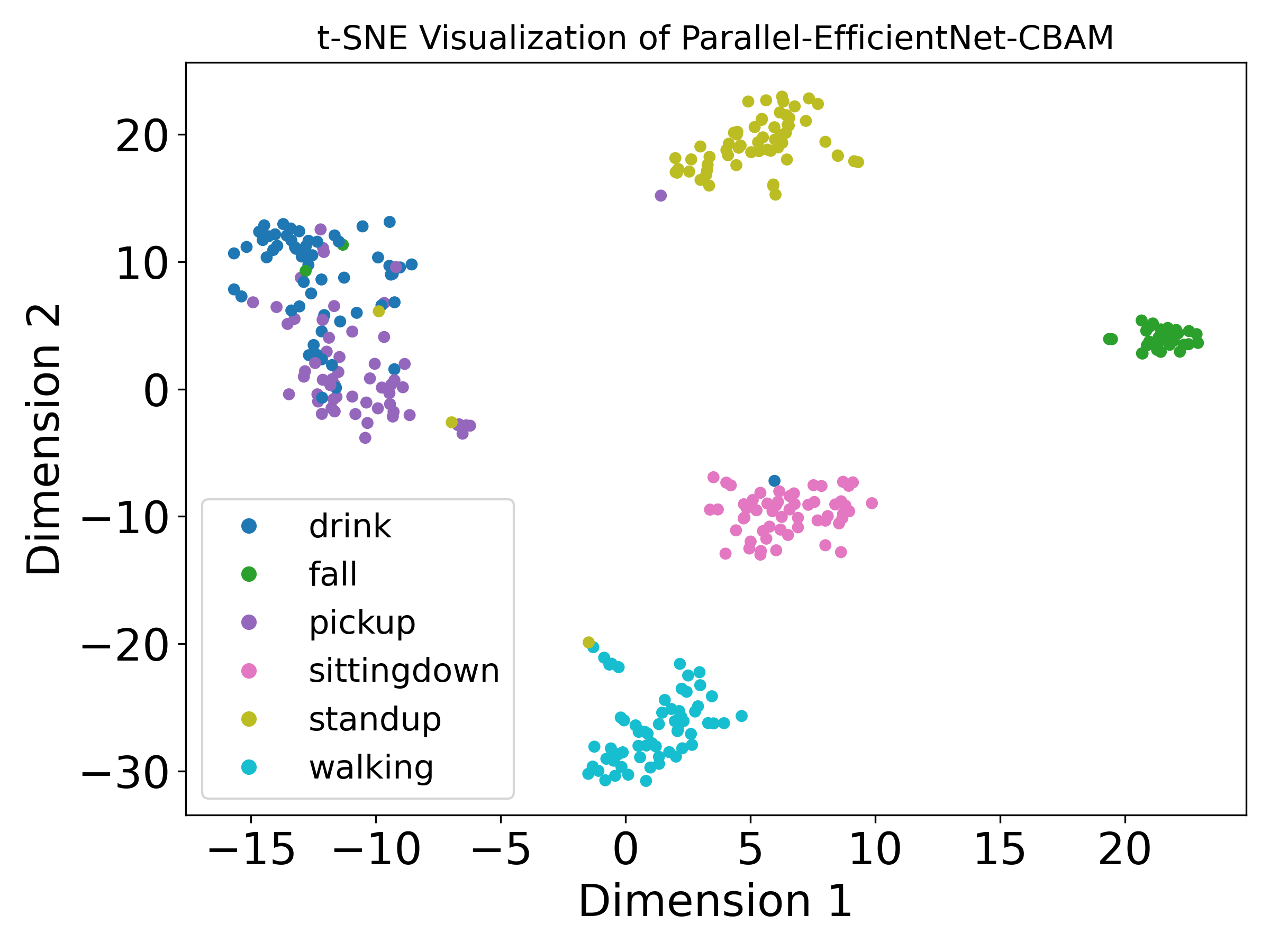}}
        \centerline{(f) \label{fig:tsne_pec}}
    \end{minipage}
    \hfill
    \begin{minipage}{0.245\linewidth}
        \centerline{\includegraphics[width=\textwidth]{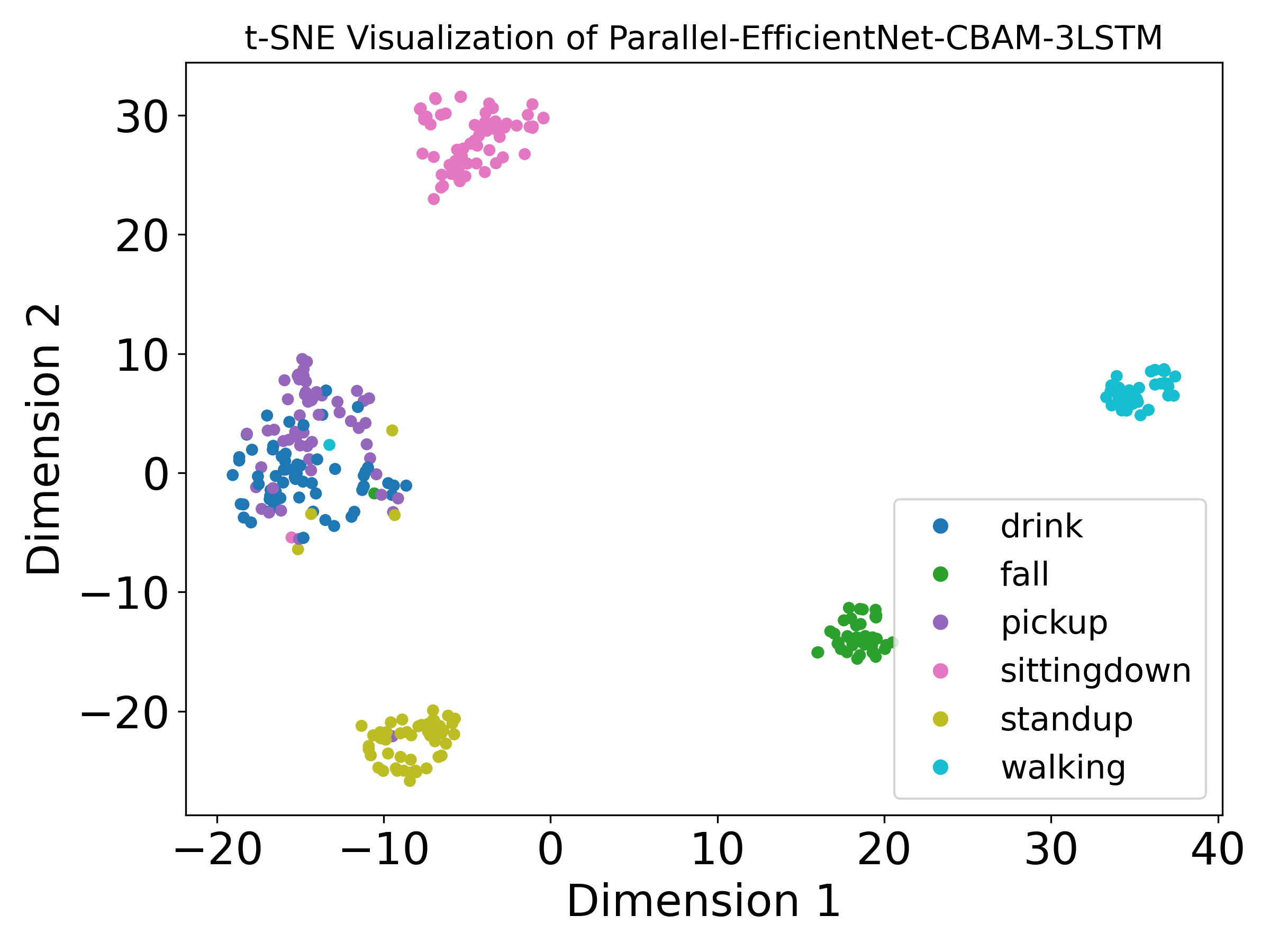}}
        \centerline{(g) \label{fig:tsne_pec3l}}
    \end{minipage}
    \hfill
    \begin{minipage}{0.245\linewidth}
        \centerline{\includegraphics[width=\textwidth]{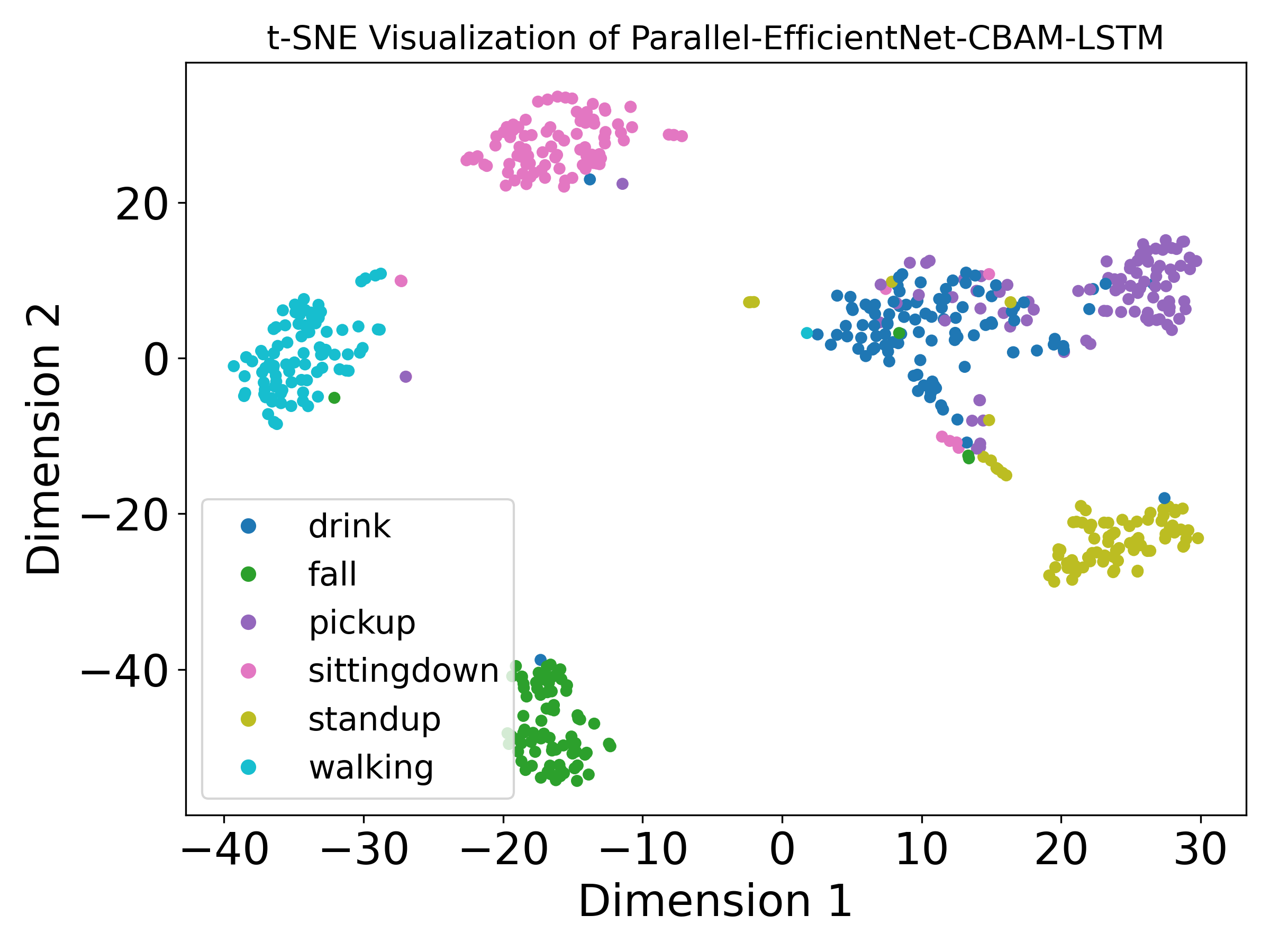}}
        \centerline{(h) \label{fig:tsne_pecl}}
    \end{minipage}
    \caption{t-SNE visualization of feature embeddings across model variants: (a) raw data; (b–d) single-branch outputs (Range-Time, Doppler-Time, Range-Doppler); (e) Parallel-EfficientNet (PE); (f) PE with CBAM (PEC); (g) PEC with LSTM on all branches; (h) our final PECL model. }
    \label{fig:tsne_all}
\end{figure*}

This performance degradation in the Range-Doppler domain led to our architectural refinement. We replaced the LSTM module with a linear layer followed by max pooling, creating the final Parallel-EfficientNet-CBAM-LSTM (PECL) architecture. Comparative analysis in Table~\ref{table:Pooling} demonstrates the superiority of max pooling over average pooling, with the former achieving 96.16\% overall accuracy compared to 78.98\% for the latter. The significant 17.18\% difference confirms that max pooling better preserves discriminative features critical for action recognition. Figures.~\ref{fig:tsne_all} (f) and (h) show the t-SNE method, which visualizes how the addition of the LSTM module helps distinguish between confusing actions. As can be seen from Figure~\ref{fig:tsne_all}(g), when all three branches employ LSTM modules, the model's classification results become confused instead. 

\begin{table}[H]
\centering
\caption{Comparison of max pooling and average pooling in the Range–Doppler branch (\%). }
\resizebox{\columnwidth}{!}{
\renewcommand{\arraystretch}{1.15}
\setlength{\tabcolsep}{4pt}
\begin{tabular}{c|cccccc|c}
\hline
\textbf{Model} & \textbf{Walk} & \textbf{Sit} & \textbf{Stand} & \textbf{Pick} & \textbf{Drink} & \textbf{Fall} & \textbf{Acc.} \\
\hline
PECL\_Avgpool & \cellcolor{gray!40}92.13 & \cellcolor{gray!60}72.95 & \cellcolor{gray!35}89.25 & \cellcolor{gray!70}54.87 & \cellcolor{gray!55}77.05 & \cellcolor{gray!10}100.00 & \cellcolor{gray!55}78.98 \\ 
\rowcolor{gray!15}
\textbf{PECL\_Maxpool} & 100.00 & 98.94 & 100.00 & 89.77 & 90.38 & 98.61 & 96.16 \\ 
\hline
\end{tabular}}
\label{table:Pooling}
\end{table}

\begin{figure}[H]
\centering
    \includegraphics[width=0.9\linewidth]{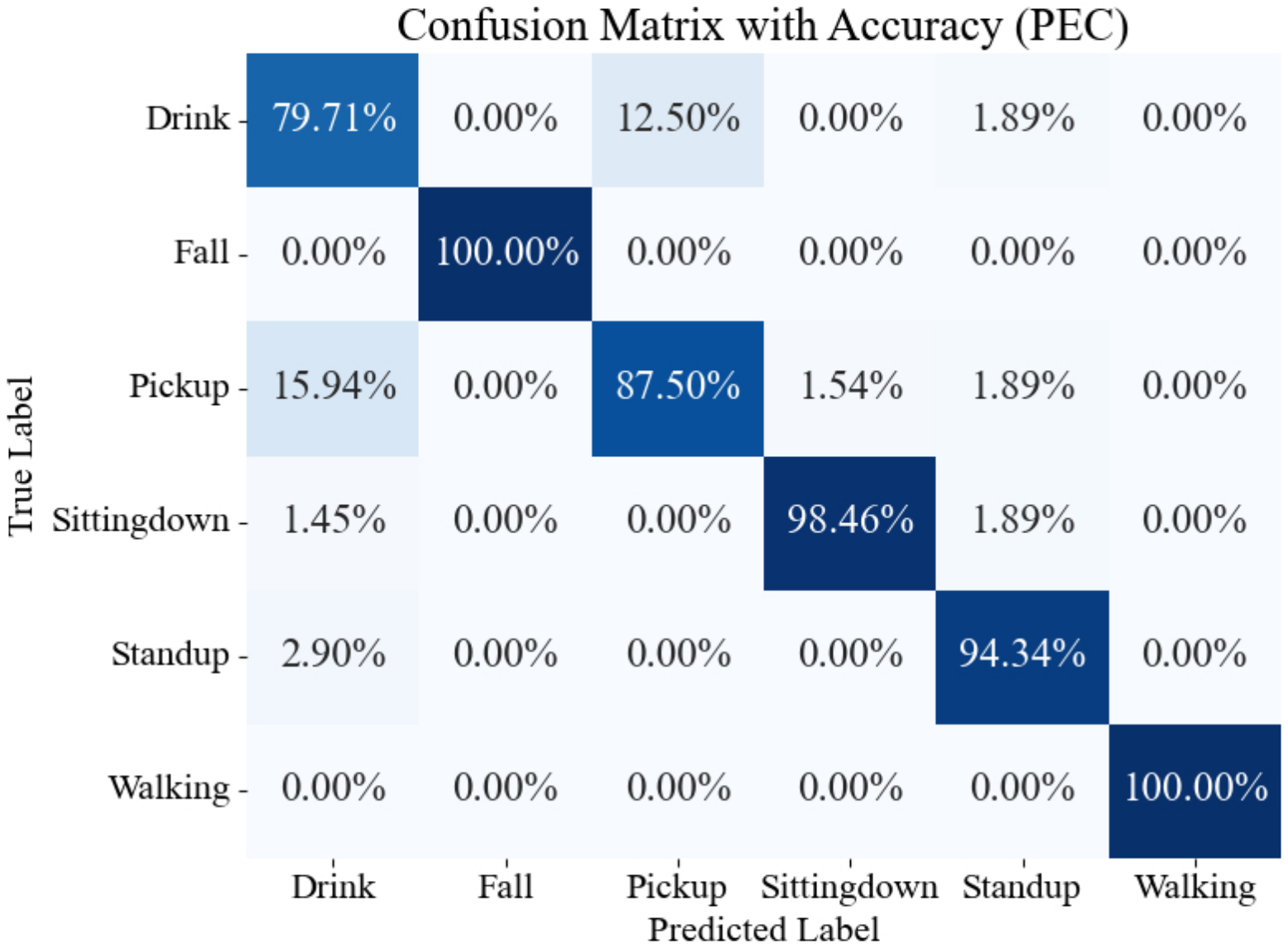}
    \centerline{(a)}

    \vspace{4pt} 

    \includegraphics[width=0.9\linewidth]{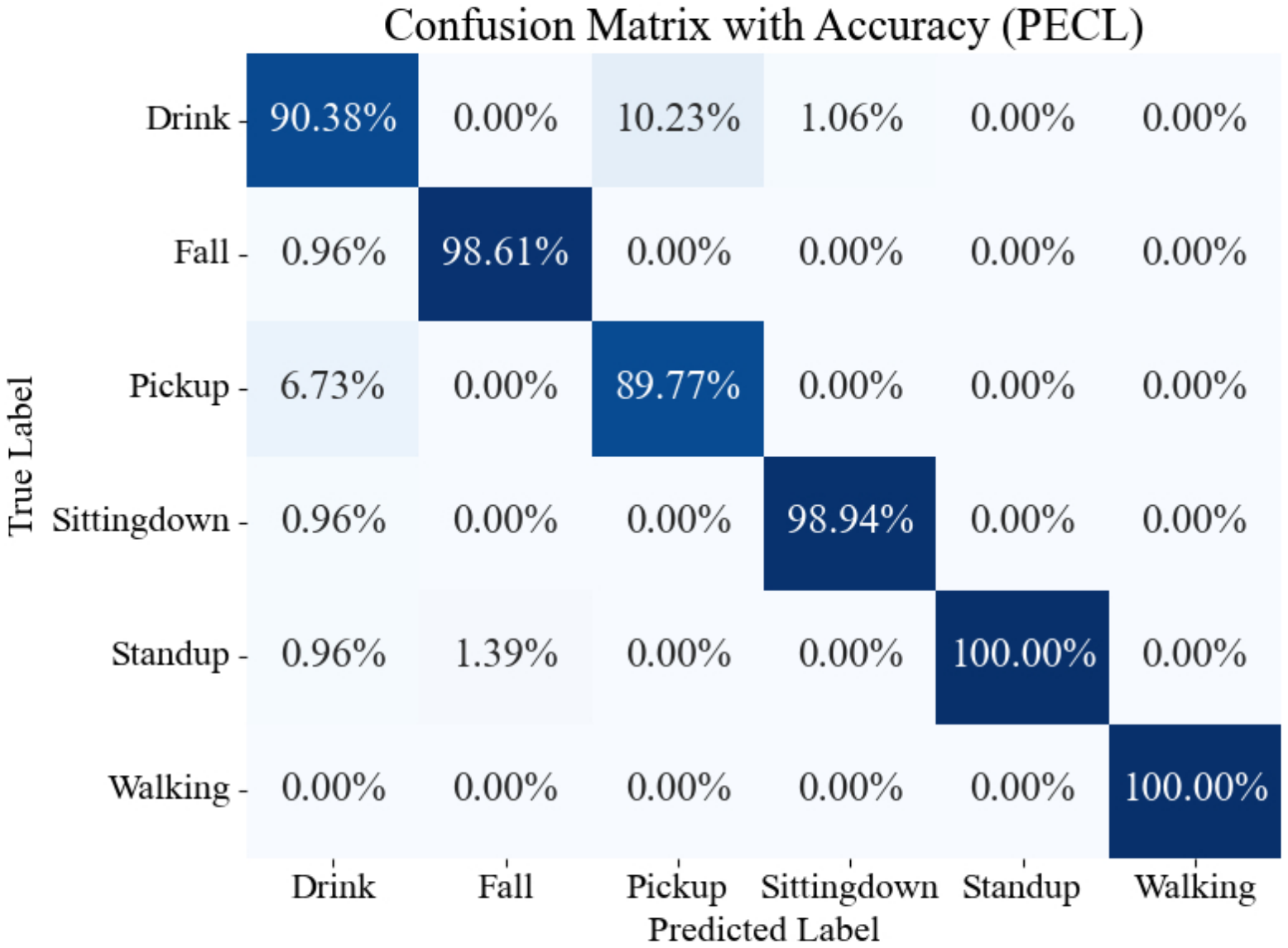}
    \centerline{(b)}

    \caption{Confusion matrices of (a) PEC and (b) PECL.}
    \label{fig:confusion_matrix}
\end{figure}

A detailed examination of the confusion matrices (Figure~\ref{fig:confusion_matrix}) highlights PECL’s enhanced ability to distinguish between the frequently confused actions “Pick Up” and “Drink.” In the baseline PEC model, these classes were misclassified at rates of 12.50\% and 15.94\%, corresponding to accuracies of 87.50\% and 79.71\%, respectively. With the integration of LSTM modules, PECL reduced these misclassification rates to 10.23\% and 6.73\%, achieving 89.77\% and 90.38\% accuracies. This improvement quantitatively confirms the effectiveness of combining temporal modeling in sequential domains with optimized spatial feature extraction in the Range–Doppler representation.

\section{Conclusion and Future Work}

\subsection{Conclusion}
This paper presented a novel architecture, Parallel-EfficientNet-CBAM-LSTM (PECL), for robust human activity recognition using radar-based sensing. The key innovation lies in the parallel extraction and fusion of three complementary spectrogram domains—Range–Time, Doppler–Time, and Range–Doppler—enabling comprehensive spatiotemporal feature modeling. PECL integrates CBAM attention modules into an EfficientNet-B0 backbone to enhance spatial feature discrimination, while LSTM units embedded in the temporal branches capture sequential dependencies critical for distinguishing subtle and visually similar actions such as “Pick Up” and “Drink.”

Experimental results on real radar datasets confirm that PECL achieves high accuracy (96.16\%) with substantially lower computational cost in terms of both parameters and FLOPs. Overall, PECL provides a balanced solution that combines high accuracy with lightweight design, demonstrating strong potential for practical radar-based HAR systems.

\subsection{Future Work}
While PECL demonstrates strong classification performance with low model complexity, several directions remain for future research.

First, although the current results are promising, they are derived from single-run experiments. Future studies should include repeated training trials with averaged metrics and standard deviations to ensure statistical robustness and reproducibility. This will provide a more comprehensive assessment of performance variance arising from random initialization and stochastic optimization.

Second, we plan to collect a proprietary radar dataset to expand the diversity and granularity of action categories. In particular, actions such as “fall” will be decomposed into subtypes (e.g., forward, backward, and lateral falls) to better capture the nuances of human activity.

Third, we aim to further optimize PECL for deployment on resource-constrained edge and mobile devices by improving the trade-off between recognition accuracy and model efficiency.

Finally, we intend to extend PECL toward multi-modal sensing by integrating RGB imagery with millimeter-wave radar data. In this framework, each modality will undergo independent feature extraction before being fused through an adaptive decision-level mechanism. This multimodal approach is expected to enhance recognition robustness while maintaining computational efficiency, paving the way for real-world intelligent monitoring applications.

\ifCLASSOPTIONcaptionsoff
  \newpage
\fi

\newpage
\bibliographystyle{IEEEtran}
\bibliography{Mybib}

\end{document}